\documentclass[useAMS,usenatbib, usegraphicx]{mn2e}
\usepackage{fancyhdr}
\usepackage{makeidx}
\usepackage{lscape}
\usepackage{hyperref}
\usepackage{wasysym}
\usepackage{longtable}
\usepackage{aas_macros}
\usepackage{txfonts}
\usepackage{times}
\usepackage{graphicx}
\usepackage{color}
\usepackage{ulem}
\newcommand\ion[2]{#1$\;${\scshape{#2}}}%                       % ion, i.e., CII = \ion{C}{ii}
\begin{document}

\title[Tangential motions in the Galactic Anti-centre with SDSS and {\it Gaia}]{The fall of the Northern Unicorn: Tangential motions in the Galactic Anti-centre with SDSS and {\it Gaia}}
\author[T.J.L. de Boer and V. Belokurov and S. E. Koposov]{T.J.L. de Boer$^{1}$\thanks{E-mail:
tdeboer@ast.cam.ac.uk} and V. Belokurov$^{1}$ and S. E. Koposov$^{1,2}$\\
$^{1}$ Institute of Astronomy, University of Cambridge, Madingley Road, Cambridge CB3 0HA, UK\\
$^2$McWilliams Center for Cosmology, Department of Physics, Carnegie Mellon University, 5000 Forbes Avenue, Pittsburgh, PA 15213, USAs
}
\date{Received ...; accepted ...}

\pagerange{\pageref{firstpage}--\pageref{lastpage}} \pubyear{2017}

\maketitle

\begin{abstract}
We present the first detailed study of the behaviour of the stellar
proper motion across the entire Galactic Anti-centre area visible in
the Sloan Digital Sky Survey data. We use recalibrated SDSS astrometry
in combination with positions from {\it Gaia} DR1 to provide
tangential motion measurements with a systematic uncertainty $<$5
kms$^{-1}$ for the Main Sequence stars at the distance of the
Monoceros Ring. We demonstrate that Monoceros members rotate around
the Galaxy with azimuthal speeds of $\sim230$ kms$^{-1}$, only
slightly lower than that of the Sun. Additionally, both vertical and
azimuthal components of their motion are shown to vary considerably
but gradually as a function of Galactic longitude and latitude. The
stellar over-density in the Anti-centre region can be split into two
components, the narrow, stream-like ACS and the smooth Ring. According
to our analysis, these two structures show very similar but clearly
distinct kinematic trends, which can be summarised as follows: the
amplitude of the velocity variation in $v_{\phi}$ and $v_z$ in the ACS
is higher compared to the Ring, whose velocity gradients appear to be
flatter. Currently, no model available can explain the entirety of the
data in this area of the sky. However, the new accurate kinematic map
introduced here should provide strong constraints on the genesis of
the Monoceros Ring and the associated sub-structure.

\end{abstract}


\label{firstpage}

\section{Introduction}
\label{introduction}
If you smash two galaxies together, the spectrum of possible outcomes
can be explored simply by dialling the mass ratio up and
down. Somewhere between the barely noticeable accretion of dwarfs onto
a giant host and the dramatic major merger events, lies the regime of
minor mergers, whose exact consequences are not easy to predict. This
is unfortunate, given that at the late stages of galaxy evolution,
i.e. z$<$1, minor mergers start to play an increasingly important
role, as the major merger rate subsides in all objects apart from the
most massive ones \citep[see e.g.][]{guo2008}. In the Milky Way, a
run-of-the-mill spiral galaxy, minor mergers can trigger some
momentous changes. These include, for example, emergence of spiral
structure \citep[see e.g.][]{mihos1994,eliche2011} as well as warping,
puffing up or even destruction of the stellar disc \citep[see
  e.g.][]{toomre1977,barnes1992,velazquez1999,steinmetz2002,benson2004}. Furthermore,
the dwarfs accreted in almost co-planar configurations can actually
contribute to the disc growth \citep[see
  e.g.][]{lake1989,abadi2003,read2008,pilepich2015,gomez2017}. Finally,
a sequence of such events can in fact lead to a complete morphological
transformation of a spiral galaxy into an elliptical one
\citep[see][]{bournaud2007}.

Locally, the most striking example of an interaction of the stellar
disc with a massive dwarf galaxy is the so called Monoceros
Ring. Analysing the first batch of the imaging data from the Sloan
Digital Sky Survey (SDSS), \citet{Newberg02} uncovered multiple
stellar over-densities outside the disc plane. These included parts of
the Sagittarius tidal stream as well as pieces of an extended and
previously unknown sub-structure in the direction of the Galactic
anti-centre. The Color-Magnitude Diagram revealed a prominent excess
of Main Sequence Turn Off (MSTO) stars with colours bluer than that of
the thick disc, located in the narrow distance range around $\sim$10
kpc from the Sun. It was hypothesised that the structure may resemble
a ring for which two formation scenarios were proposed: debris form a
disrupted dwarf galaxy and an ``even thicker disc''. Shortly after,
\citet{Ibata03} carried out their own CMD analysis of the SDSS
photometry which they supplemented by the imaging data collected with
the Isaac Newton Telescope on La Palma. They confirmed the ring-like
appearance of the Monoceros structure and put forward four hypotheses
to explain its emergence: i) a pattern the Galactic disc would develop
over time, similar to a warp, and possibly a result of an interaction
with accreted satellites, ii) tidal debris from a disrupted dwarf
galaxy, iii) part of the outer spiral arm, and, finally iv) stellar
disc resonances induced by e.g. Milky Way's bar.

\begin{figure}
\centering
\includegraphics[angle=0, width=0.495\textwidth]{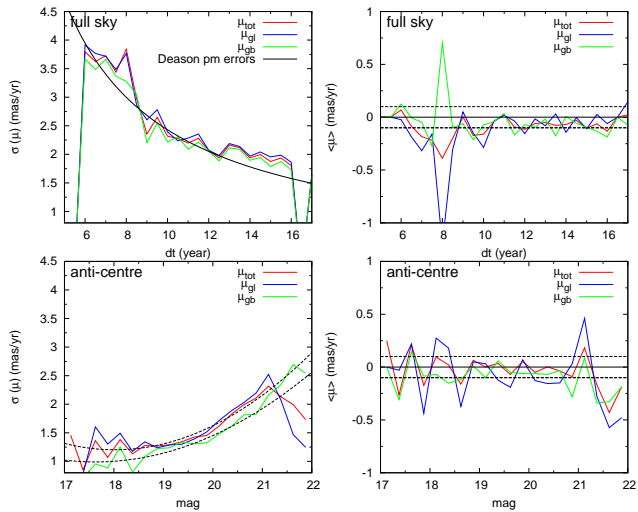}
\caption{Distribution of the proper motions of spectroscopic QSOs~\citep{Paris17} in the SDSS-{\it Gaia} sample, for total proper motion ([$\mu_{l}$,$\mu_{b}$]) as well as individual components. {\it Top left:} Dispersion of the proper motion components as a function of time baseline for the full-sky sample, showing that the full-sky sample is consistent with uncertainties as determined in~\citet{Deason17}. {\it Top right:} Median of the proper motions for the full-sky sample as function of time baseline showing a systematic uncertainty of $\approx 0.1$\,mas/yr. {\it Bottom left:} Dispersion of the proper motion components in the Galactic anti-centre as a function of g-band magnitude. The anti-centre sample has a limited range of time in the flat long baseline regime, and shows little variation as function of baseline. Black lines indicate fits to the proper motion dispersion in $\mu_{l}$ and $\mu_{b}$ components, to be used in constructing Monoceros models. {\it Bottom right:} Median of the proper motions for the full-sky sample as function of magnitude showing a systematic uncertainty of $\approx 0.1$\,mas/yr. \label{Mono_errors}}
\end{figure}

The follow-up work by \citet{Yanny03} helped to nudge the opinion of
the community towards the tidal debris origin of the Monoceros
Ring. The conclusion seemed inevitable as \citet{Yanny03} measured low
stellar velocity dispersion values - between 20 kms$^{-1}$ and 30
kms$^{-1}$ - for several sight-lines through the ring, only slightly
higher than that obtained for the Sgr stream stars. They also
estimated the metallicity of the constituent MSTO stars to be of order
of [Fe/H]=-1.6, much lower than that of either the thin or the thick
disc. Thus started the hunt for the Ring's progenitor. Following the
successful identification of the Ring with the 2MASS M giants by
\citet{rocha2003}, \citet{martin2004a} detected a stellar over-density
whose position and shape appeared to match the description of the
presumed progenitor. Dubbed Canis Major, the object proved to be
notoriously difficult to probe given its location very close to the
Galactic plane (at $b=-8^{\circ}$). Depressingly, as subsequent
studies piled up, the nature of the object in the constellation of
Canis Major became less clear. On the one hand, the structural
parameters of CMa - as measured e.g. by \citet{bella2006} - looked
very similar to those of a tidally disrupted dwarf galaxy. Even more
convincingly, using the MMT's Hectospec, \citet{martin2005} obtained a
very small velocity dispersion of $\sim$10 kms$^{-1}$ for stars in
CMa. On the other hand, \citet{Momany04}, with the help of the 2MASS
Red Clump and Red Giant stars, demonstrated that CMa could in fact be
a part of the Galactic warp. Another demonstration of the similarity
of the stellar populations in CMa and in the Galactic disc was
provided by the study of \citet{mateu2009} who found no evidence for
the presence of a significant number of RR Lyrae stars in the vicinity
of the object.

To add to the confusion, \citet{Crane03} measured much higher average
metallicity of [Fe/H]=-0.4 for the M giant stars in the Monoceros Ring
as compared to the value [Fe/H]=-1.6 reported by \citet{Yanny03}. Not
only such high metallicity agrees well with both thin and thick disc
populations, additionally, the line-of-sight velocities in their
sample are broadly consistent with the hypothesis in which the member
stars are on circular orbits with v$_{\phi}=220$\,kms$^{-1}$. This
kinematic pattern was confirmed in the subsequent spectroscopic
studies of \citet{conn2005} and \citet{martin2006}. A different
approach was taken by \citet{Ivezic08}, who instead of relying on
spectroscopy, utilised the power of the SDSS imaging to constrain the
metallicity and the kinematics of the Monoceros Ring. Using multi-band
photometry, they produced a photometric metallicity distribution
function for the MS stars, and found that the typical [Fe/H] value for
the Monoceros members is around -0.95 with a narrow dispersion of 0.14
dex. Using the proper motions of stars around $l=180^{\circ}$, they
concluded that Monoceros may rotate even faster than the LSR, with
speeds up to 270 kms$^{-1}$.

\begin{figure*}
\centering
\includegraphics[angle=0, width=0.95\textwidth]{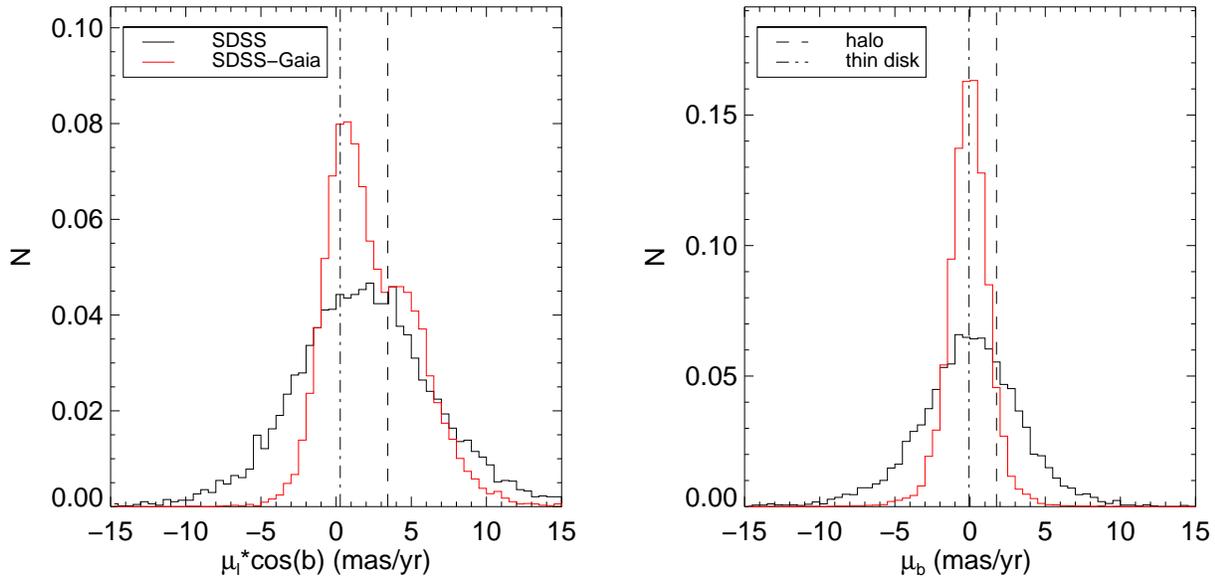}
\caption{Normalised proper motion histogram of the ACS stream as in Figure 18 of~\citet{Li12}, but in Galactic longitude/latitude coordinates. Black histograms show the proper motions of stars as measured in SDSS DR9, while red histograms show proper motions from our new SDSS-{\it Gaia} sample. The selected sample covers the area $160<l<190$\,deg, $29<b<38$\,deg and selects faint turn-off stars $19<g<20$, $0.2<(g-r)<0.3$ and $(u-g)>0.4$. The expected proper motion of a non-rotating halo from the Galaxia model is indicated by the dashed line~\citep{Sharma11}, while the dash-dotted line shows $220$\,km/s disk rotation at a distance of 10 kpc. Note that the proper motion distribution in $\mu_{\mathrm{l}}$ is clearly resolved in two components, one of which is consistent with non-rotating MW halo and one with a disk-like rotation signal. \label{Lidist_comp}}
\end{figure*}

The spectroscopic portrait of the Ring started to come into focus when
based on low-resolution MMT Hectospec spectroscopy of Main Sequence
stars,~\citet{Meisner12} found a metallicity of $[{\rm Fe/H}]\approx-1.0$\,dex, slightly more metal-poor than the MW
thick disk. The stars were also found to be under-abundant in [Ca/Fe]
compared to the MW thick disk and kinematically cold with
$\sigma_{rad}\approx30$\,kms$^{-1}$. This metallicity measurement is
in striking disagreement with both \citet{Yanny03} and
\citet{Crane03}, and matches perfectly the results of
\citet{Ivezic08}. There are good reasons to trust the 2012 study more
than either of the two early estimates. First, unlike \citet{Yanny03},
\citet{Meisner12} actually measure [Fe/H] directly from the
spectra. Second, in contrast to \citet{Crane03}, they measure the
abundance of the MS stars, i.e. the most prominent Monoceros Ring
tracer population. Note, however, that the different metallicity
measurements discussed above could actually be reconciled if the
Ring's MDF was sufficiently broad. This would make it look more like
the Sgr dwarf debris, but also more similar to the disc. In the
opinion of \citet{Meisner12} however, the abundance pattern was more
consistent with the dwarf debris origin, based on the metallicity
dispersion appropriate for the progenitor's luminosity $M_{\rm v}$ and
the measured [Ca/Fe] deficiency.

As a result of the intense scrutiny described above, two competing
hypotheses came to the forefront: one which invoked the production of
the Ring via the disruption of a yet-to-be-discovered dwarf galaxy (or
possibly CMa) and another which posited that Monoceros was not a
structure on its own but rather a perturbation of the Galactic
disc. In 2006, an interesting line of enquiry highlighted a new
possibility. According to \citet{Grillmair06}, in the direction of the
Galactic anti-centre, there existed both a broad diffuse component as
well as at least two narrower, stream-like structure (dubbed ACS, or
Anti-centre Stream) with very similar CMDs. They also discovered a
fainter and shorter structure dubbed Eastern Banded Structure
(EBS). They argued that the smooth component represented dwarf galaxy
debris and the narrow bands were streams from sub-systems,
e.g. globular clusters. Note that according to the re-analysis of EBS
by \citet{Grillmair11}, it was unlikely to be related to the rest of
the Monoceros. Nevertheless, the discovery of several components to
the Ring, each with different structural properties, indicated that
Monoceros might not be a single entity with a simple genesis scenario,
but rather a superposition of many. Most recently, \citet{Li12} used
SEGUE spectroscopy of the Galactic anti-centre to determine a
metallicity of $[{\rm Fe/H]}=-1.0$\, dex for the ACS region above
$b>25$ deg and $[{\rm Fe/H}]=-0.8$\,dex for the lower latitude Ring
region, thus providing further supporting evidence that the two
closely situated - and in fact overlapping - structures have somewhat
distinct properties.

The idea that the Galactic anti-centre is significantly more messy
than previously envisaged is supported by the detection of the
Southern Galactic counterpart to the Monoceros Ring, the so-called
Tri-And structure. \citet{rocha2004} used 2MASS M giants to trace a
broad arc of stars under the Galactic disc, but at larger distances
compared to Monoceros, i.e. $>15$ kpc from the Sun. They also
collected velocities of some of the stars in Tri-And which they used
to argue that the structure does not rotate as fast as the thin
disc. Over the years, pieces of Tri-And kept popping up
serendipitously in different surveys \citep[see
  e.g.][]{majewski2004,martin2007,martin2014}. This motivated
\citet{deason_triand} to compile a large sample of photometric and
spectroscopic detections from a variety surveys across the Tri-And
area in an attempt to consolidate the view of its properties. Their
study established the following. First and foremost, similar to
Monoceros in the Galactic North, Tri-And appears to be littered with
sub-structure, which included narrow stellar streams as well as
satellites. They also demonstrated that along several selected SDSS
sight-lines, the transition between the disc and Tri-And was not
smooth and that most likely, Tri-And stars rotated slower than the
Galactic thick disc.

Could the Monoceros Ring and Tri-And be related? Certainly, this thought
has crossed many a mind, and this is exactly the idea entertained by
\citet{rocha2004} who pointed out the seemingly smooth transitions
between the two structures in the plane of line-of-sight velocity as a
function of Galactocentric longitude. While it is tempting to link the
two structures, even if the are located at different distances from
the Galactic centre, there may exist noticeable differences in the
spectroscopy of their stars. For example, \citet{chou2010} found the
Monoceros MDF to be similar to that of the Sgr dwarf or perhaps even
the LMC. However, in the subsequent study of the Southern Galactic
hemisphere, they detected enough differences between Monoceros and
Tri-And \citep[][]{chou2011}. These inferences appear at odds with the
hypothesis put forward by \citet{Xu15}. By mapping out the asymmetries
between the North and South, they argue that the various structures in
the anti-centre direction form one wave-like pattern which shows up as
stellar density enhancement or depletion above and below the disc
plane as a function of heliocentric distance. Note however, that it
remains unclear whether the disc oscillation can explain the entirety
of the tangled sub-structure visible in the panoramic maps constructed
with PS1 catalogs as presented in \citet{Slater14}.

Many of the observational studies above attempted to compare the
measured properties of the Monoceros Ring to the results of numerical
simulations, no matter how scarce. The first such simulation was
provided by \citet{helmi2003} who considered the disruption of a
satellite on a nearly circular orbit. They pointed out two distinct
outcomes, depending on the age of the accretion event: narrow arc-like
streams or, alternatively, shells of stellar debris. While the work by
\citet{helmi2003} was only the proof of principle,
\citet{Penarrubia05} assembled the first comprehensive model of the
Monoceros Ring, which explained many of the observations available at
the time. The two studies described above can be contrasted with the
analysis carried out by \citet{Kazantzidis08} in which a relatively
low-mass ($\sim 10^{10}$M$_{\odot}$) CDM sub-structure was used to
perturb the disc through a series of interactions starting some 8 Gyr
ago. Superficially, both the dwarf disruption and the disc
perturbation models gave a satisfactory explanation of the
Ring. However, one tricky question remained unanswered: how does one
position a satellite on such a low-radius circular orbit?
Accordingly, \citet{younger2008} demonstrated that
cosmologically-motivated eccentric orbits can also produce cold
ring-like structures in the disc, with stars reaching considerable
heights above the plane. With the possible mechanism established, two
plausible perturbers were considered: the Sgr dwarf galaxy
\citep[e.g][]{Purcell11} and the LMC \citep[e.g.][]{laporte2016}. Sgr
in particularly appears to be a good candidate, as its orbit crosses
the disc at the Galactocentric distance similar to that of the
Monoceros Ring. Typically, in the simulations of a massive dwarf
in-fall, the authors would rely on dynamical friction to circularise
the orbit. However, \citet{michel2011} pointed out a different
mechanism for the Sgr orbit circularisation: a three-body encounter.

Given the early measurements of the radial velocity and the proper
motion, as well as the variety of other circumstantial links with the
Galactic disc, the bulk of the motion of the Moncoeros Ring stars is
likely in the plane perpendicular to the line-of-sight, at least in
the direction of the Galactic anti-centre. Motivated by this, here we
attempt to provide insights into the nature of Monoceros by using a
combination of SDSS and {\it Gaia} data (referred to as SDSS-{\it
  Gaia} from now on) to investigate the proper motion distribution of
the Monoceros Ring across the Galactic anti-centre
region~\citep{Ahn14,GAIAmain1,GAIAmain2}. The combination of the SDSS
and {\it Gaia} source catalogs allows us to determine proper motions
for the faint, blue Main Sequence~(MS) stars - the best available
tracer of Ring to date - with an unprecedented accuracy. Compared to
most other dataset of similar depth and area, our SDSS-{\it Gaia}
proper motion sample suffers no noticeable systematic errors and
deteriorates very little as a function of apparent magnitude. We fit a
simple cylinder-like model to the data and use synthetic MW models to
account for the effects of contamination by the thick disk and
halo. In this way, we aim to extract information about the 3D velocity
components of Monoceros at different Galactic latitudes, with the hope
to procure new constraints on the nature of this enigmatic structure.

This Paper is organised as follows: in Section~\ref{data} we present
the proper motion dataset, in Section~\ref{selection} we describe the
selection of a clean sample of Monoceros stars within the SDSS
footprint, after which the proper motions as a function Galactic
latitude are presented and interpreted in
Section~\ref{propermotions}. Section~\ref{Monomodels} describes the
construction of models for the Monoceros Ring and shows the proper
motion trends as a function of the different
parameters. Section~\ref{Monofitting} gives the model fits to the data
and the velocities for the Monoceros components inferred from
those. This is followed by a full two dimensional fit in
Section~\ref{spatfits}. Finally, Section~\ref{conclusions} discusses
the results and their implications for the origin of the Monoceros
Ring.

\section{SDSS-{\it Gaia} proper motions}
\label{data}

To study the proper motion of the Monoceros Ring, we make use of SDSS Data Release 10~\citep{Ahn14} as well as data obtained by the {\it Gaia} satellite~\citep{GAIAmain1,GAIAmain2}. In particular, we use the {\it Gaia} source catalog, which provides accurate positions of nearly one billion stars~\citep{Lindegren16}. Comparison between the positions in both catalogs yields proper motions across a large area of the sky, more precise than previously available. However, as the SDSS positions are based on astrometric solution tied to the UCAC survey~\citep{Zacharias00},  any systematics inherent to UCAC as well as low density of UCAC calibrators leads to systematics on the proper motions through the naive SDSS-Gaia crossmatch. To remove these systematics and increase the precision of proper motions, Koposov et al (in prep) recalibrate the astrometry of SDSS DR10 based on the excellent astrometry of the Gaia source catalog. We refer to their paper and~\citet{Deason17} for more details of the recalibration procedure. In short, the astrometric calibration adopted in SDSS is rerun following the method of~\citet{Pier03}, but instead of the UCAC-4 catalog, the Gaia source catalog is used as astrometric calibrators. Following this recalibration, the sources in SDSS and the Gaia catalogs are cross-matched using nearest-neighbor matching with a 10 arcsecond aperture, resulting in proper motions for the majority $r\lesssim 20$ sources in the entire SDSS footprint with a baseline of $\approx$10 years in the Northern hemisphere and $\approx$5 years in the South.

To determine the uncertainties on the proper motion measurement of SDSS-{\it Gaia}, we make use of the SDSS DR12 sample of spectroscopically confirmed quasars~\citep{Paris17}. Under the safe assumption that the quasars do not move, the median and dispersion in the proper motion of the quasar sample determines the systematics and uncertainty on the positions of Gaia and SDSS measurements, as shown in Figure~\ref{Mono_errors}. First, we consider the full-sky sample of quasars and compute median and dispersion (calculated as 1.48 times the median absolute deviation) of the total proper motion $(\mu_{l},\mu_{b})$ as well as individual components, following~\citet{Deason17}. The top panels of Figure~\ref{Mono_errors} show that the proper motion uncertainty depends critically on the time baseline between the SDSS and {\it Gaia} observation, and is consistent with the dependence as a function of baseline fitted by \citet{Deason17}. Note also that the uncertainties of the $\mu_{l}$ and $\mu_{b}$ components are very similar. We compute a median offset of $\approx 0.1$\,mas/yr, similar to~\citet{Deason17}. Next, we constrain our sample to the Galactic anti-centre where we will study the Monoceros features. The anti-centre sample covers the long baseline regime of 10$-$15 year separation between SDSS and {\it Gaia} observations. Therefore, little variation of proper motion uncertainties is seen as function of time. Instead, in the bottom panels of Figure~\ref{Mono_errors} we show the median and dispersion of the QSOs as a function of the g-band magnitudes. A clear trend as function of magnitude is visible due to the lower precision of faint objects. Notably, the uncertainties of the $\mu_{b}$ component are clearly smaller than the uncertainties of $\mu_{l}$, due to the alignment of SDSS scans with the $\mu_{b}$ direction in the anti-centre. The black lines indicate quadratic polynomial fits to the proper motion dispersion in $\mu_{l}$ and $\mu_{b}$ components, which will be used in constructing models of Monoceros across the anti-centre. Best-fit curves in the anti-centre are given by the functions $\sigma_{\mu_l} = 36.04 - 3.87 g + 0.11 g^{2}$ and $\sigma_{\mu_b} = 26.50 - 2.89 g + 0.08 g^{2}$.

To test the quality and validity of the SDSS-{\it Gaia} proper motions, we compare our results to those of a series of measurement in different regions of the Monoceros stream. Using kinematics of stars in Kapteyn's selected area 76 ($l=209.3$\,deg, $b=26.4$\,deg),~\citet{Carlin10} selected a clean sample of 31 Monoceros stars from which they measure a mean $\mu_{\alpha}\,\cos(\delta)=-1.20\pm0.34$\,mas/yr, $\mu_{\delta}=-0.78\pm0.36$\,mas/yr. From our catalog, we select the same spatial region and apply a similar colour and magnitude selection to extract Monoceros stars. From our resulting sample of $\approx$100 stars, we measure a peak and standard deviation of proper motion of $\mu_{\alpha}\cos{\delta}=-0.81\pm 0.98$\,mas/yr, $\mu_\delta = -1.23 \pm 2.02$\,mas/yr which is marginally different but consistent within the errors. Other measurements of the proper motion in the same region have resulted in values of $\mu_{\alpha}\cos{\delta}=0.67\pm0.81$\, mas/yr, $\mu_\delta=0.73\pm0.80$\,mas/yr~\citep{Grillmair08} and $\mu_{\alpha}\cos{\delta}=-0.55\pm0.40$\, mas/yr, $\mu_\delta=-0.58\pm0.33$ mas/yr~\citep{Li12}, showing there is significant variation (i.e. proper motions of almost opposite signs) between different works. Typical measurement uncertainties on individual points is 1$-$2 mas/yr for~\citet{Carlin10} up to $\approx$4 mas/yr for~\citet{Grillmair08, Li12}, not taken into account when deriving mean values. We conclude our measured values are consistent with other works within the measurement errors. 

Another test of the accuracy of our proper motions is done by comparing proper motion  distributions of a large sample of Monoceros stars as defined in~\citet{Li12}. Following their work, we select a sample of stars in the Galactic anti-centre region between $160<l<190$ deg and $29<b<38$ deg, adopting a simple colour and magnitude cut to select blue MS stars~($19<g<20$, $0.2<(g-r)<0.3$ and $(u-g)>0.4$). Subsequently, we create histograms of the proper motion distribution, similar to Figure~18 of~\citet{Li12} but in Galactic coordinates, shown for our sample in Figure~\ref{Lidist_comp}. A comparison between both panels shows that the distribution of our sample is much narrower than that of SDSS DR9, indicating that the proper motion errors in our SDSS-{\it Gaia} catalog are significantly smaller. Furthermore, the proper motion distribution in $\mu_{\mathrm{l}}$ as measured by our catalogue is clearly resolved into two components, where one is consistent with a halo population~(dashed line, from the Galaxia model) and the other with a disk-like population at 10 kpc~(dash-dotted line) as expected for Monoceros. This conclusively shows that our proper motion sample has smaller uncertainties, and much better able to distinguish different kinematic populations.

\section{Sample selection}\label{selection}
\begin{figure}
\centering
\includegraphics[angle=0, width=0.495\textwidth]{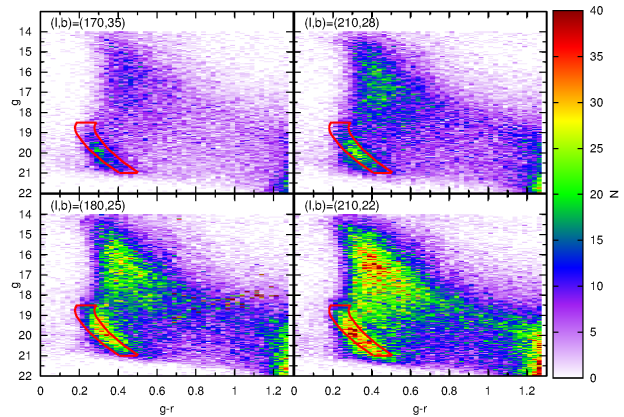}
\caption{The colour-magnitude diagram of four fields across the anti-centre region of SDSS, displaying the prominent Monoceros MS sequence. An isochrone with the age of $8$\,Gyr, $[{\rm Fe/H}]=-1.0$\,dex at a distance of $10$\,kpc and  offset by $0.05$ in colour redward and blueward is overlaid to illustrate the selection of our Monoceros sample. We only consider stars with magnitudes between $19<g<21$ to avoid highly contaminated CMD regions. \label{mono_CMD}}
\end{figure}

The Monoceros Ring is most prominent in the SDSS data in the Northern hemisphere around the Galactic anti-centre~\citep{Xu15}. Across this region, the Monoceros stars are visible as a prominent MS which spans a small range of distances around $10$\,kpc from the Sun~\citep{Li12}. To obtain a clean sample of Monoceros stars, we adopt an isochrone appropriate for Monoceros (with an age of $8$\,Gyr and $[{\rm Fe/H}]=-1.0$\,dex) and select all stars within $0.05$ of the MS between $19<g<21$. Figure~\ref{mono_CMD} shows the CMD of four fields in the anti-centre region, with two isochrones offset from each other in colour by $0.05$ in magnitude to highlight our adopted selection. This Figure shows that the prominent Monoceros MS sequence is indeed well contained within our selection in several areas of the sky occupied by the Monoceros structure.

\begin{figure*}
\centering
\includegraphics[angle=0, width=0.975\textwidth]{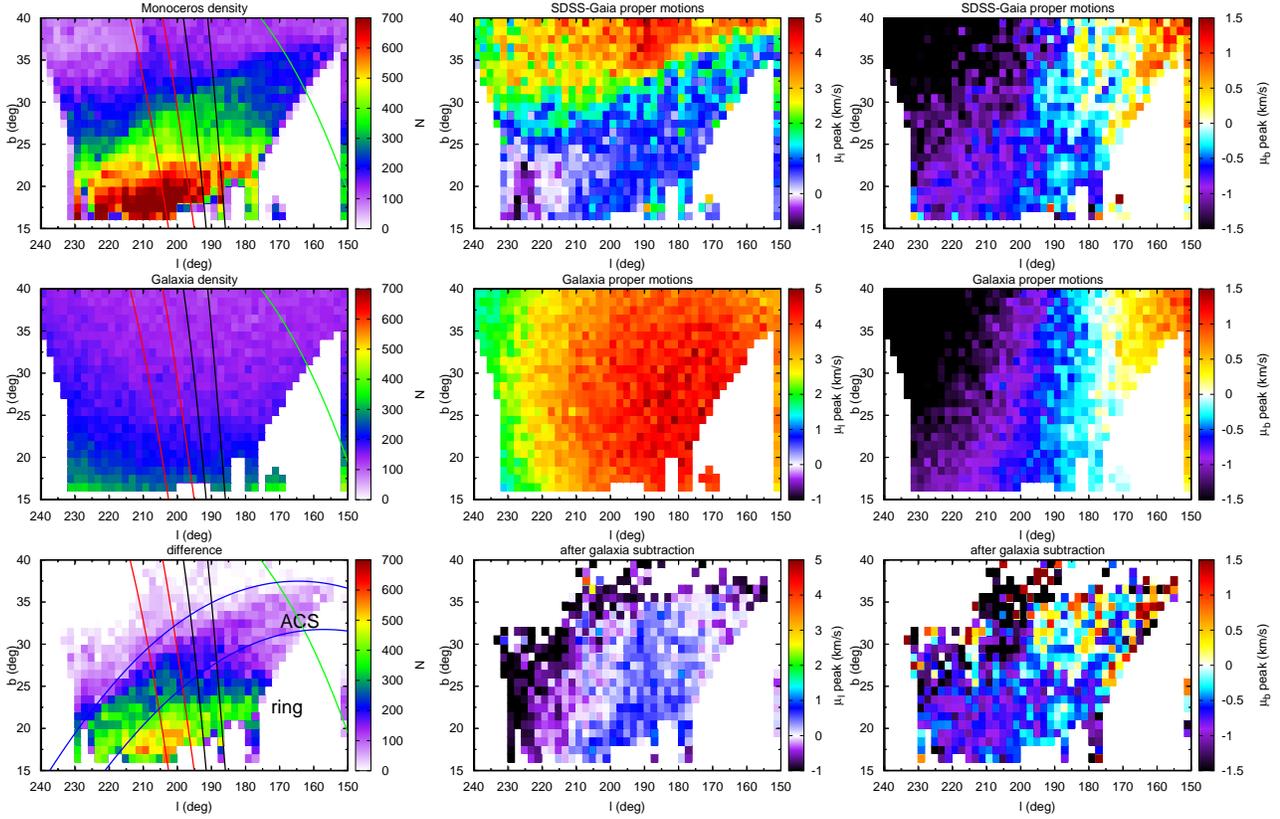}
\caption{Density and proper motions of our Monoceros sample in comparison to the Galaxia model, as function of Galactic longitude and latitude using 2$\times$1 deg bins. {\it Left panels:} Density maps showing the Monoceros sample in SDSS-{\it Gaia} on the top, the Galaxia model (using the same colour-magnitude selections) in the middle and the difference in the bottom. Monoceros is visible as several high density shells running diagonally from the bottom left to the top right. Lines indicate the trajectory of known streams in this part of the sky, including the Sagittarius bright stream~(red lines), Sagittarius faint stream~(black lines) and the Orphan stream in green~\citep{Belokurov072, Koposov12}. Blue lines show great circles indicating the area of sky associated with the ACS stream, with the area below the blue lines occupied by the Monoceros Ring. {\it Middle panels:} The mode of the proper motions in the $\mu_{l}$ direction for the Monoceros sample in the top panel, and the Galaxia model in the middle panel. The bottom panel shows the mode of the proper motion determined after subtracting the MW model from the Monoceros sample. {\it Right panels:} The mode of the proper motions in the $\mu_{b}$ direction for the Monoceros (top), and Galaxia (middle) samples, as well as the mode after subtracting both distributions (bottom). \label{Mono_densitypm}}
\end{figure*}

The left panels of Figure~\ref{Mono_densitypm} show a density plot of stars within our Monoceros selection in Galactic coordinates, using 2$\times$1 deg bins. For comparison, we also show a density plot of Milky Way stars generated using the Galaxia model of the MW~\citep{Sharma11}, which contains a detailed spatial and kinematic model of the different Milky Way components. We generate a Galaxia realisation in small bins of Galactic longitude and latitude respectively, and use the synthetic magnitudes to apply the same colour-magnitude selection as for the real data (after applying photometric scatter due to magnitude dependent errors). Finally, the bottom left panel of Figure~\ref{Mono_densitypm} shows the Galaxia-subtracted density of our Monoceros sample.

\begin{table*}
\caption[]{Proper motion parameters of Monoceros as a function of Galactic latitude, given as the mode and dispersion (calculated as 1.48 times the median absolute deviation) of the histograms shown in Figure~\ref{Mono_pmhists}. Also shown is the percentage of stars within our CMD selection belonging to different components of the MW according to the Galaxia model, indicating what the main contaminating MW population is. Population fractions are display as frac$_{thin disk}$/frac$_{thick disk}$/frac$_{halo}$. Note that above b$\approx$35 deg, there is very little Monoceros stars left and the signal is dominated by MW halo contamination.}
\begin{center}
%\resizebox{0.48\textwidth}{!}{

\begin{tabular}{cccccr}
\hline\hline
b &  $\left<{\mu_{l}}\right>$ & $\sigma_{\mu_{l}}$ & $\left<{\mu_{b}}\right>$ & $\sigma_{\mu_{b}}$ &\multicolumn{1}{c}{f$_{pop}$}  \\
deg & mas/yr & mas/yr &mas/yr & mas/yr & \\
\hline
 16$-$18 & -0.06$\pm$0.02 & 2.31$\pm$0.01 & -0.98$\pm$0.01 & 1.78$\pm$0.01 & 6.6/45.0/48.3 \\
 18$-$20 &  0.07$\pm$0.02 & 2.20$\pm$0.01 & -0.99$\pm$0.01 & 1.67$\pm$0.01 & 2.6/38.9/58.5 \\
 20$-$22 & -0.01$\pm$0.02 & 2.23$\pm$0.01 & -0.97$\pm$0.01 & 1.69$\pm$0.01 & 1.3/32.3/66.4 \\
 22$-$24 &  0.12$\pm$0.02 & 2.32$\pm$0.01 & -0.97$\pm$0.01 & 1.77$\pm$0.01 & 0.6/26.7/72.8 \\
 24$-$26 &  0.27$\pm$0.02 & 2.52$\pm$0.02 & -0.96$\pm$0.01 & 1.84$\pm$0.01 & 0.2/21.9/77.8 \\
 26$-$28 &  0.58$\pm$0.02 & 2.69$\pm$0.02 & -0.91$\pm$0.01 & 1.89$\pm$0.01 & 0.1/18.1/81.8 \\
 28$-$30 &  0.82$\pm$0.03 & 2.95$\pm$0.02 & -0.96$\pm$0.02 & 1.96$\pm$0.01 & 0.1/14.7/85.2 \\
 30$-$32 &  1.22$\pm$0.03 & 3.04$\pm$0.02 & -0.93$\pm$0.02 & 2.07$\pm$0.01 & $<$0.1/11.4/88.5 \\
 32$-$34 &  1.50$\pm$0.03 & 3.21$\pm$0.02 & -0.95$\pm$0.02 & 2.10$\pm$0.01 & $<$0.1/09.7/90.3 \\
 34$-$36 &  1.82$\pm$0.03 & 3.33$\pm$0.02 & -0.89$\pm$0.02 & 2.34$\pm$0.02 & $<$0.1/08.1/91.9 \\
 36$-$38 &  3.00$\pm$0.04 & 3.47$\pm$0.03 & -0.83$\pm$0.02 & 2.56$\pm$0.02 & $<$0.1/06.7/93.3 \\
 38$-$40 &  2.85$\pm$0.04 & 3.32$\pm$0.03 & -0.98$\pm$0.02 & 2.59$\pm$0.02 & $<$0.1/05.6/94.4 \\
\hline 
\end{tabular}
%}
\end{center}
\label{Monopars}
\end{table*}

The main Monoceros Ring is visible as an arc spanning from the bottom right to the top left. Below the prominent arc, several other arcing features are visible, all of which are part of the Monoceros Ring and anti-centre stream. Besides the Monoceros Ring, this region of the sky is also host to the Sagittarius streams, which are outlined by the red and blue lines for the bright and faint components respectively~\citep{Koposov12}. Furthermore, the trajectory of the Orphan stream is indicated by the green line crossing over the Monoceros feature at low longitude~\citep{Belokurov072}. However, these intersecting streams are located at distances much larger than the Monoceros features ($22.5$\,kpc for Orphan and $25-35$\,kpc for Sagittarius), and are not expected to induce significant contamination of our sample~\citep{Casey13,Belokurov14}.

\section{Proper motion properties of the Monoceros Ring}
\label{propermotions}

We now investigate the proper motion properties of the Monoceros sample and make a comparison to the proper motions expected for the Milky Way in this region of the sky. We once again use the Galaxia model to generate samples of synthetic Milky Way stars and use the Cartesian 3D positions and velocities to compute proper motions. In the middle and right panels of Figure~\ref{Mono_densitypm} we show the proper motion of our sample in small spatial bins of 2$\times$1 deg, for respectively the Galactic longitude and latitude components. In each pixel, we show the mode of the proper motion distribution to allow for asymmetry or skew in the data. Top panels show the observed proper motions, while the middle panels show proper motions obtained from the Galaxia model. Under the assumption that the Galaxia model correctly reproduces the MW distribution in both kinematics and stellar density, we also subtract proper motion histograms of Galaxia (satisfying the same selection criteria as the data) from the observed proper motion histograms to obtain the MW corrected proper motion distribution. Therefore, we show the mode of those proper motion maps in the bottom panels of Figure~\ref{Mono_densitypm}, but only for pixels where the spatial density after subtraction is greater than the Poisson uncertainty of the MW model to avoid residual contamination.

\begin{figure*}
\centering
\includegraphics[angle=0, width=0.95\textwidth]{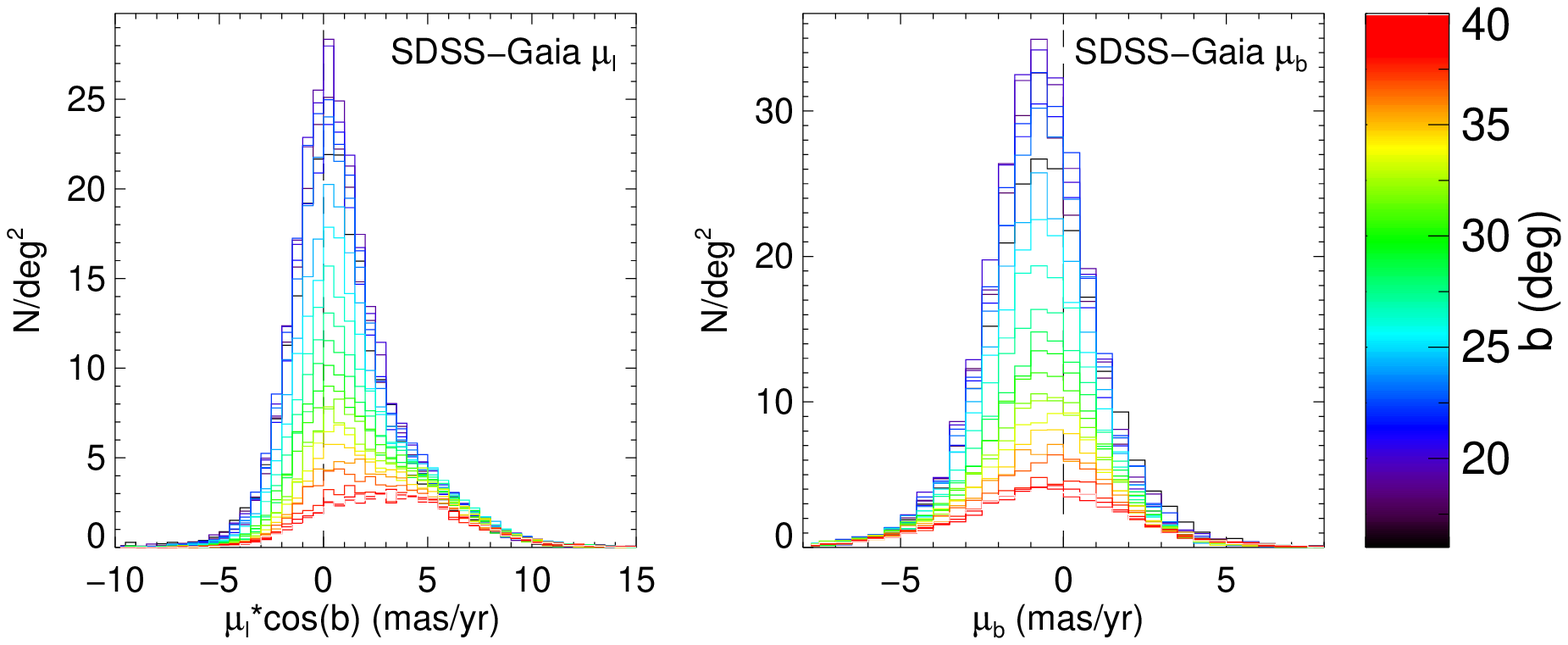}
\includegraphics[angle=0, width=0.95\textwidth]{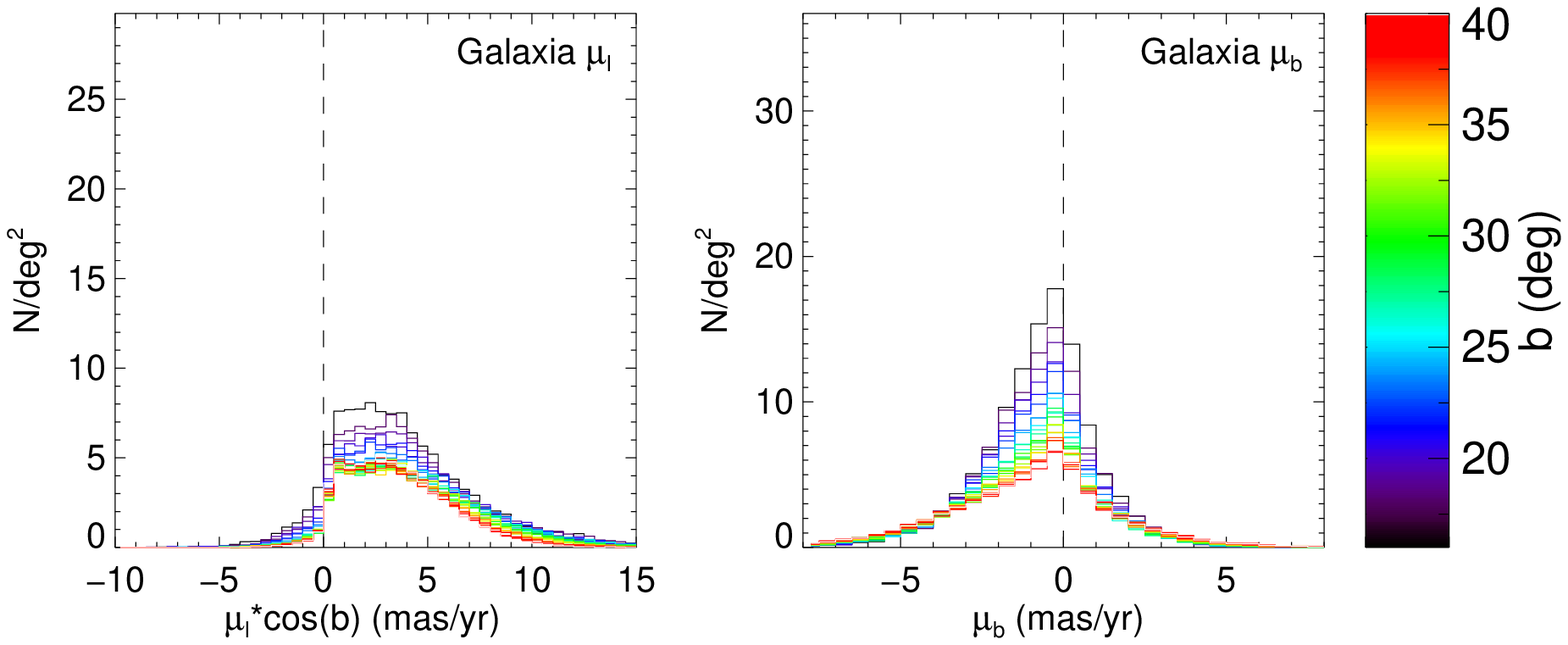}
\caption{Comparison of proper motion distribution in Galactic coordinates across the range of Galactic latitude studied. Upper panels show the proper motions for the isochrone selected Monoceros sample, while lower panels show the proper motions for the Galaxia model.  \label{Mono_pmhists}}
\end{figure*}

The Galaxia model is dominated by halo stars at the latitudes under investigation, which is reflected in the proper motion distributions in the middle panels of Figure~\ref{Mono_densitypm} . We see a clear signal due to reflex motion of the non-rotating Galactic halo, which causes high proper motion in $\mu_{l}$ and zero proper motion in $\mu_{b}$ proper motion centred on the Galactic anti-centre. The proper motions distribution is symmetric around the anti-centre as the solar reflex effect is simply dependent on angle in the Galactic plane. The distinct maximum in the middle panel of Figure~\ref{Mono_densitypm} is not centred on $b=0$ due to the presence of the thick disk, which contributes non-negligibly at lower latitudes, thereby lowering the average proper motion. Furthermore, the $\mu_{b}$ proper motion is symmetric, but the location of the $\mu_{b}=0$ contour is slightly offset from the anti-centre due to Solar motion and position in the Galaxy. 

The distribution of observed $\mu_{l}$ proper motions shows a clear signature of the Monoceros ring that correlates with the stellar density also shown in Figure~\ref{Mono_densitypm}. Stars with latitudes higher than the sharp ACS feature are characterised by high $\mu_{l}$ indicative of MW halo populations, while stars at lower latitude are clearly different from what is expected in the Galaxia model. The feature designated as the EBS by~\citet{Grillmair06} is also apparent as a sequence of lower $\mu_{l}$ stars around $(l,b)=(225,30)$. In the proper motions along Galactic latitudes (right panels of Figur~\ref{Mono_densitypm}), deviations from the MW model are less striking, due to the smaller velocities in the vertical direction. 

The Galaxia-subtracted proper motion maps in the bottom panels of Figure~\ref{Mono_densitypm} show more details of the Monoceros signal, highlighted by the selected colour range. The diagonal Monoceros sequences in $\mu_{l}$ have proper motions well below that of the halo but larger than zero, indicating a velocity smaller than the Sun. A clearer signal is also seen in the $\mu_{b}$ distribution, with Monoceros sequences corresponding to slightly negative proper motion, indicating a negative velocity compared to the Sun. We note that for high latitudes the proper motions are still consistent with the MW halo, since no strong Monoceros component is present.

The most striking change in the proper motions of Monoceros is seen as function of Galactic latitude. Therefore, we also construct the projected proper motion histograms in slices of constant Galactic latitude, both for the SDSS-{\it Gaia} dataset as well as for the Milky Way model. Figure~\ref{Mono_pmhists} shows the histograms of the observed Monoceros sample in the top panels, with colours representing the Galactic latitude. As a comparison, the bottom panels of the Figure show the histograms obtained from the Galaxia models covering the same range in latitudes. 

Figure~\ref{Mono_pmhists} shows a clear trend as function of Galactic latitude for the Monoceros sample, in both components of proper motion. For low latitudes, Monoceros shows a narrow but asymmetric distribution of $\mu_{l}$  with a peak around zero, while for increasing latitude the $\mu_{l}$ distribution becomes broader and shifts to higher values. This behaviour is not reproduced in the halo dominated Galaxia model distributions, which shows very little change in peak location and width. Monoceros also shows slight but noticeable trends in the distribution of $\mu_{b}$ where the peak shifts from $\approx-$1 to almost zero when moving from low to high latitude. To further quantify the behaviour of Monoceros, we determine the mode and dispersion (calculated as 1.48 times the median absolute deviation) of the proper motion distribution in $2$ deg wide bins of Galactic latitude. Parameters obtained for Monoceros are given in Table~\ref{Monopars}, showing a smooth and consistent change in the parameters of Monoceros with latitude, in both proper motion components. Furthermore, Table~\ref{Monopars} also shows what percentage of the Monoceros-like selection from Galaxia model belongs to each of the MW components: thin disk, thick disk and halo. As expected, the contribution of MW thin disk is very low, given the sampled range of latitudes as well as the CMD cuts adopted for our sample. Contribution from the halo in the CMD selection rises from $\approx 70$ percent at low latitudes to $94$ percent at our highest latitude. Note that above $b\approx35 $\,deg, the signal in our sample is dominated by MW halo contamination, due to the almost complete absence of Monoceros stars.

\begin{figure}
\centering
\includegraphics[angle=0, width=0.495\textwidth]{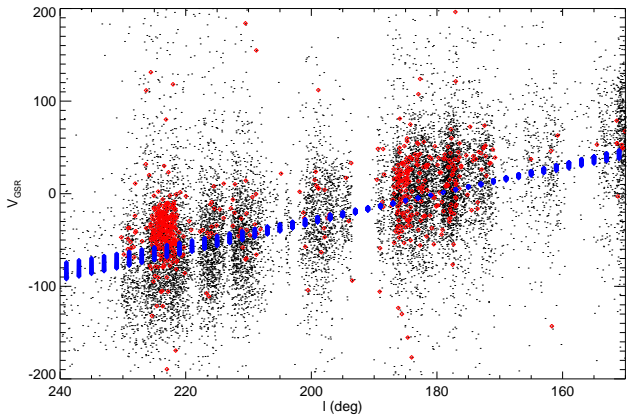}
\caption{Radial velocity (in the Galactic standard of rest frame) for spectroscopically confirmed, metal-rich~($-1.2<[{\rm Fe/H}]<-0.5$) dwarfs in the Galactic anti-centre, as a function of longitude~(black points). Red points indicate the subset of Monoceros main sequence stars classified as dwarfs~(log g$>$3.5) and satisfying the colour-magnitude mask described in section~\ref{selection}. The blue points show the expected velocities of stars in a disk rotating with $220$\,km/s, under the assumption that the stars are $10$\,kpc from the Sun~\citep{Gilmore00,Li12}. \label{Mono_specvel}}
\end{figure}

\section{Monoceros models}
\label{Monomodels}

To model the proper motion histograms shown in Figure~\ref{Mono_pmhists} and learn more about the velocities, we assume that Monoceros can be represented by a cylinder centred on the Galactic centre with a relatively small range in distance~(i.e. a torus). Based on measurements by~\citet{Xu15}, Monoceros has a well defined distance of $10$\,kpc in the direction of Galactic anti-centre, with only small variation of $\approx$10 percent. Therefore, we construct models by generating particles assuming each Monoceros star is on a ring-like shell aligned with the Galactic plane and centred on the MW centre, without any tilt or warp. We assume a distance of $10$\,kpc toward the anti-centre and adopt $8.33$\,kpc for the distance of the Sun from the Galactic center~\citep{Gillessen09}. We sample a range in $\phi$ and z coordinates sufficient to cover the anti-centre region. Subsequently, we adopt a large grid of possible velocities in $\phi$ and z coordinates and generate proper motion histograms for each combination, convolving with the spatial density distribution of the Monoceros sample. For all models, we add Solar reflex motion to the model velocities using value for the solar motion of v$_{\phi,\odot}=239.5$\, km/s and (U, V, W)$_{\odot}$ = (11.1, 12.24, 7.25)\, km/s from~\citet{Reid04,Schonrich10}, to construct models directly comparable to observations. Using this grid, we can efficiently construct models with a large range of velocities and velocity dispersions, to be matched against the observed proper motion histograms. 

To take into account the observed uncertainty on the proper motions of Monoceros stars, we use the relations derived from Figure~\ref{Mono_errors} for the $\mu_{l}$ and $\mu_{b}$ components as a function of magnitude. For each model, we use the g-band magnitudes to determine the average proper motion uncertainty and perturb the proper motion of each star by Gaussian errors, thereby creating models that can be compared directly to the observed data. The models produced in this way are relatively simple, and lack several thing that thin disk models typically include, such as asymmetric drift, warp and flare. However, we did not attempt to include these features given that they are completely unconstrained at these Galactocentric distances, but instead choose to study the overall properties of Monoceros using these simple models.

\begin{figure}
\centering
\includegraphics[angle=0, width=0.495\textwidth]{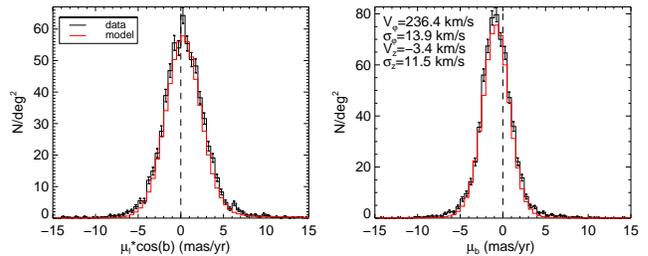}
\caption{Proper motion distributions for a sample of thin disk stars covering a Galactic longitude between $200<l<205$ deg and latitude between $5<b<10$ deg. The sample has been selected using an isochrone mask with an age of 5 Gyr and [Fe/H]=$-$0.5 dex at a distance of 3 kpc to obtain a clean sample of thin disk stars. Black lines indicate the histogram of observed data along with Poisson error bars, while the red histograms show the best-fit cylinder model (in red). Best-fit parameters are also shown in the plot. \label{test_thindisk}}
\end{figure}

For radial velocities adopted in the models, we can make use of constraints from spectroscopic observations of the SEGUE survey in the anti-centre. Following~\citet{Li12}, we select all stars in the anti-centre classified as dwarfs~(${\rm log g} > 3.5$) with metallicities consistent with Monoceros~($-1.2<[{\rm Fe/H}]<-0.5$) and investigate radial velocity variation as a function of Galactic longitude~(see Figure~\ref{Mono_specvel}). We clean our sample further by imposing the same colour-magnitude filter as described in section~\ref{selection}, resulting in the red points in Figure~\ref{Mono_specvel}. The measured velocities are a decent match to the expected velocities of stars in a disk rotating at $220$\,km/s~(shown as blue points in Figure~\ref{Mono_specvel}), under the assumption that the stars are $10$\,kpc from the Sun~\citep{Gilmore00,Li12}. Therefore, when constructing our models, we assume that the radial velocities of Monoceros satisfy this relation as a function of longitude. 

\begin{figure*}
\centering
\includegraphics[angle=0, width=0.495\textwidth]{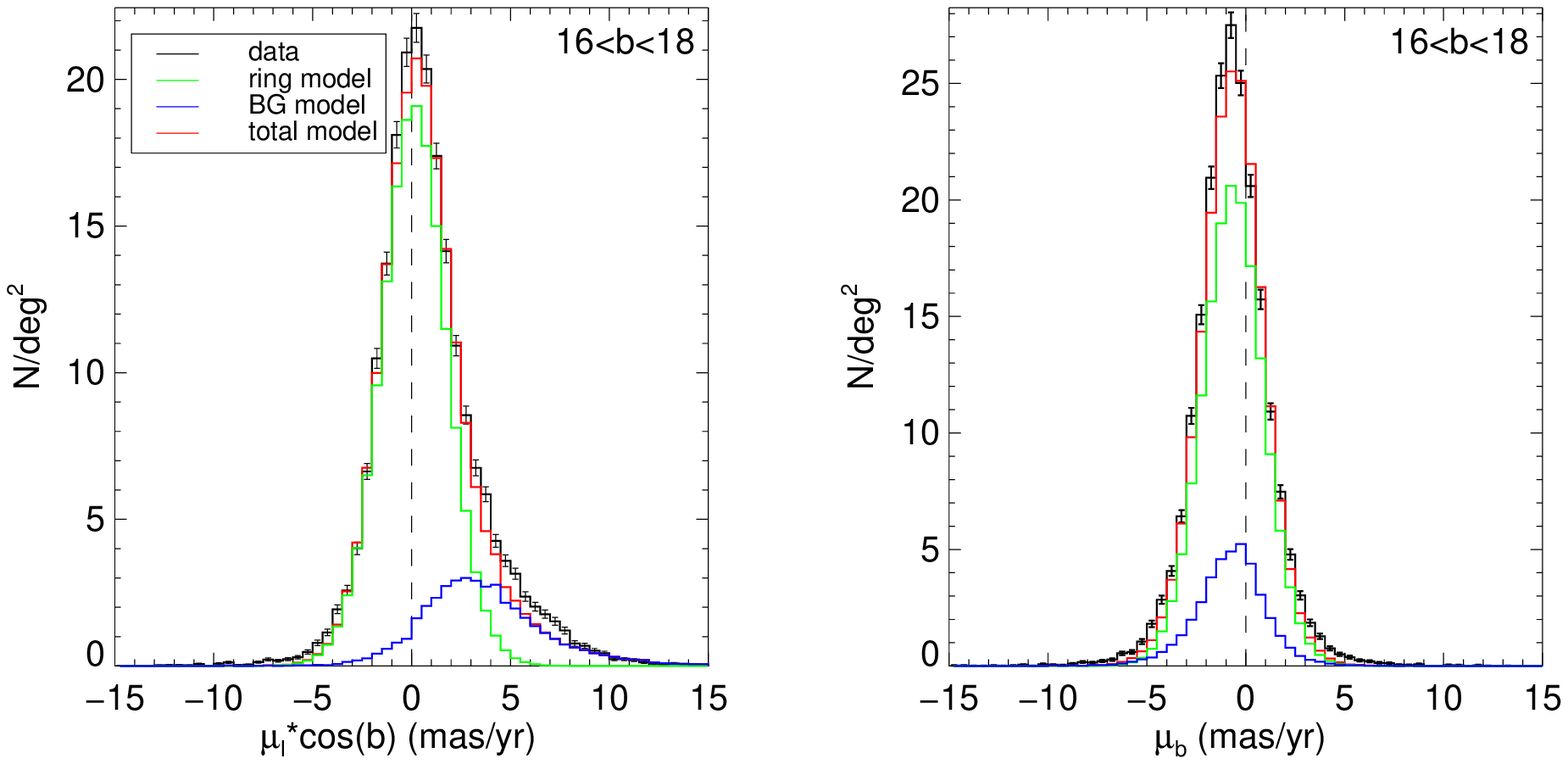}
\includegraphics[angle=0, width=0.495\textwidth]{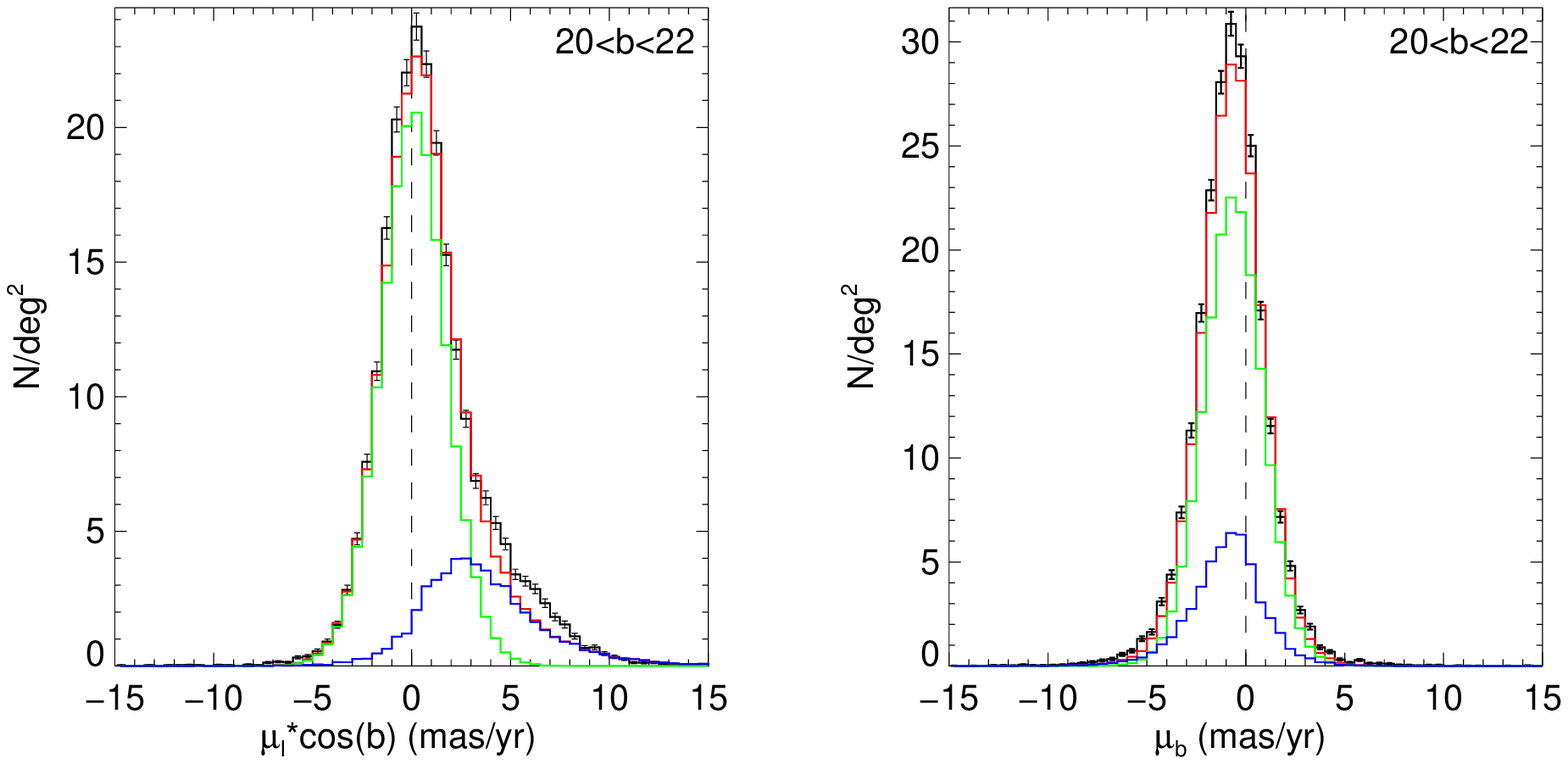}
\includegraphics[angle=0, width=0.495\textwidth]{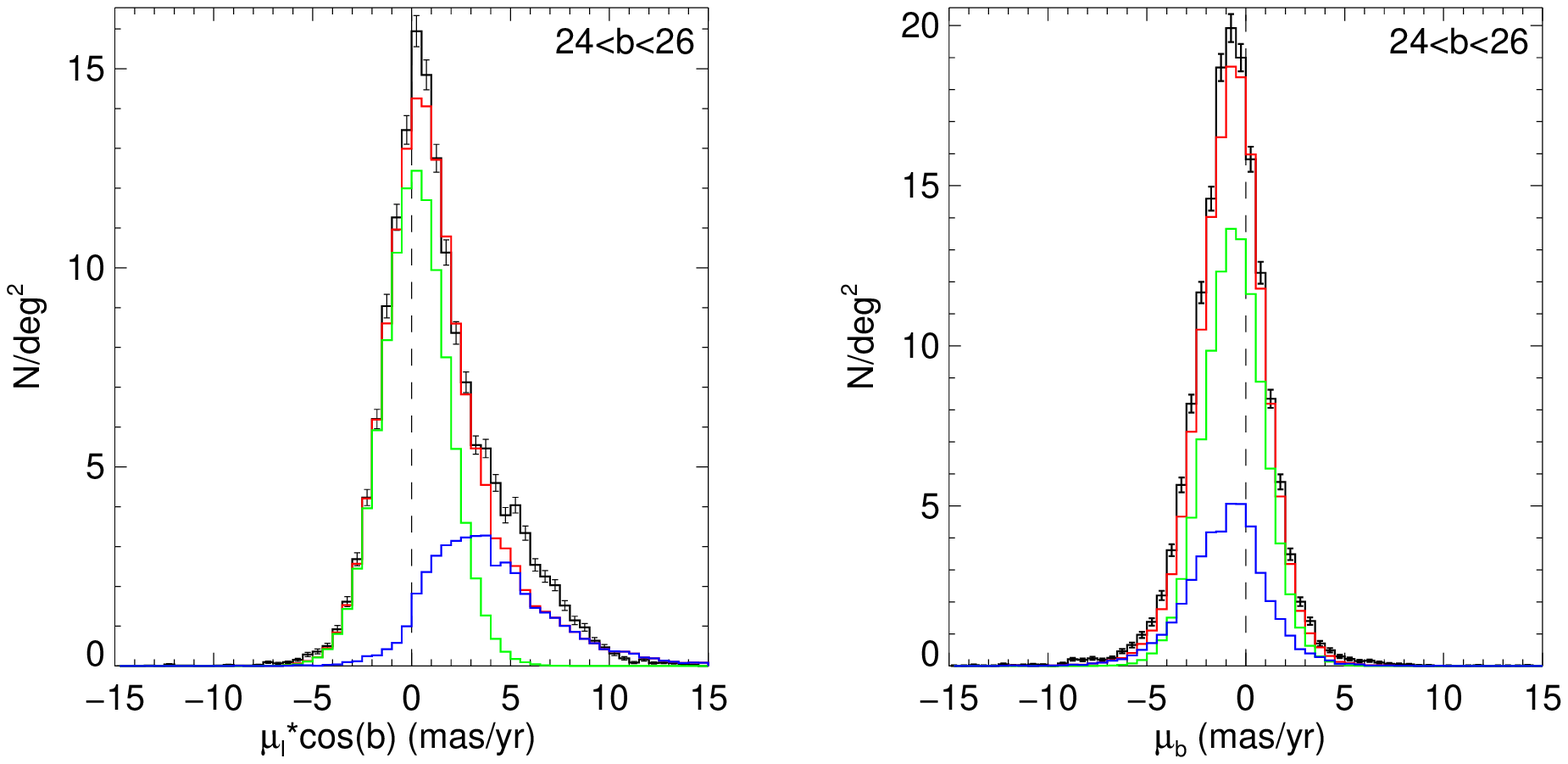}
\includegraphics[angle=0, width=0.495\textwidth]{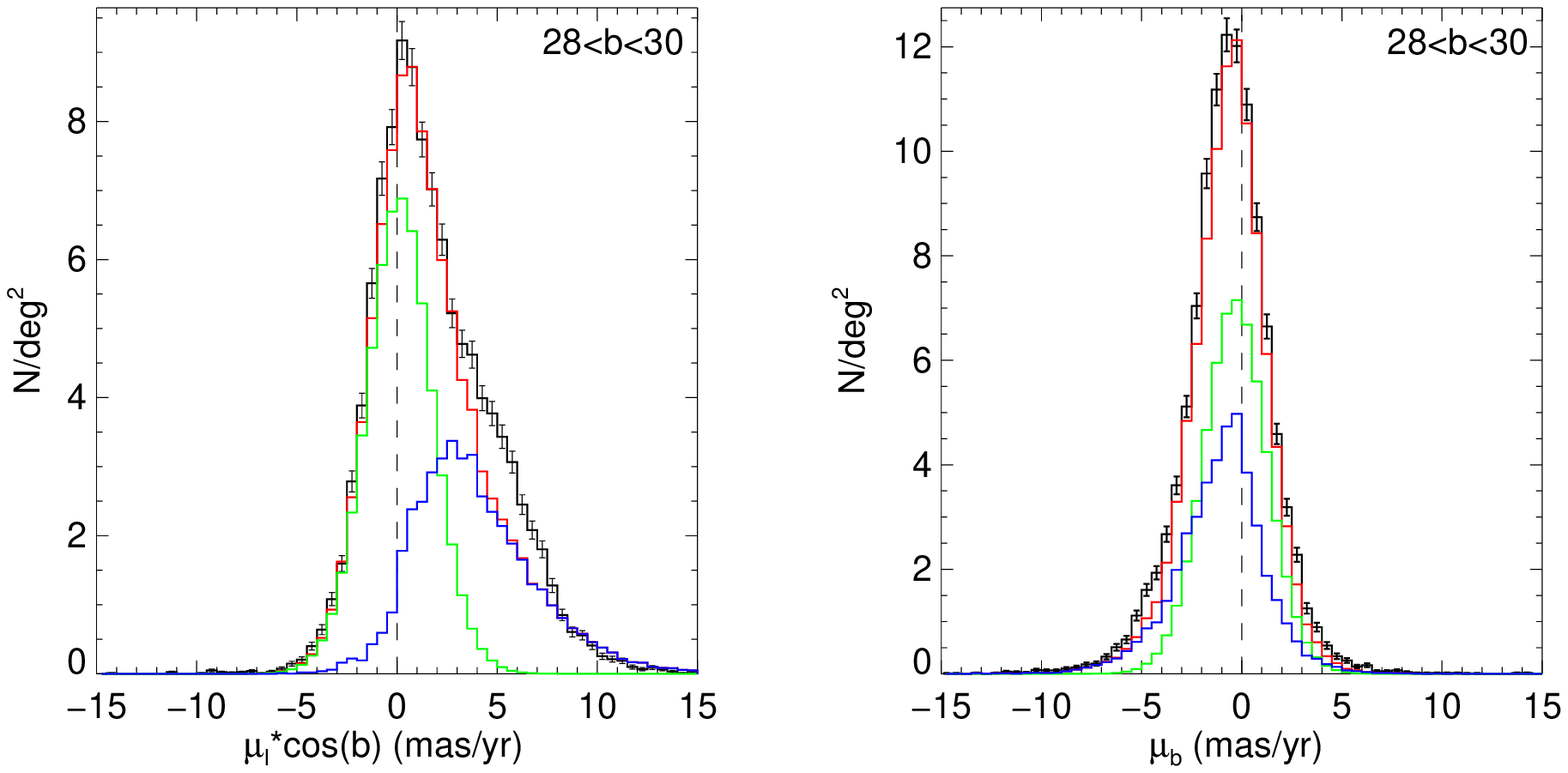}
\includegraphics[angle=0, width=0.495\textwidth]{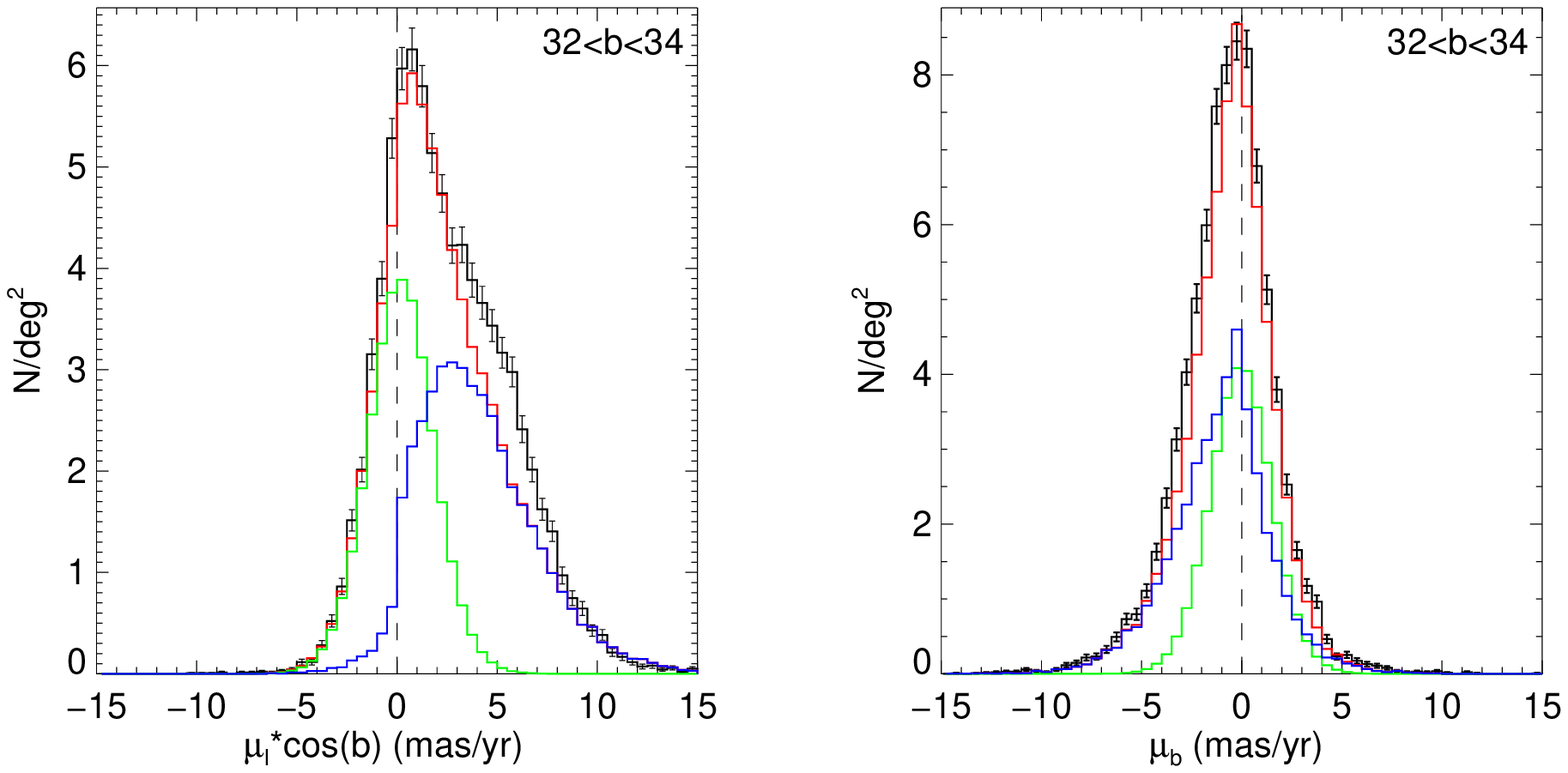}
\includegraphics[angle=0, width=0.495\textwidth]{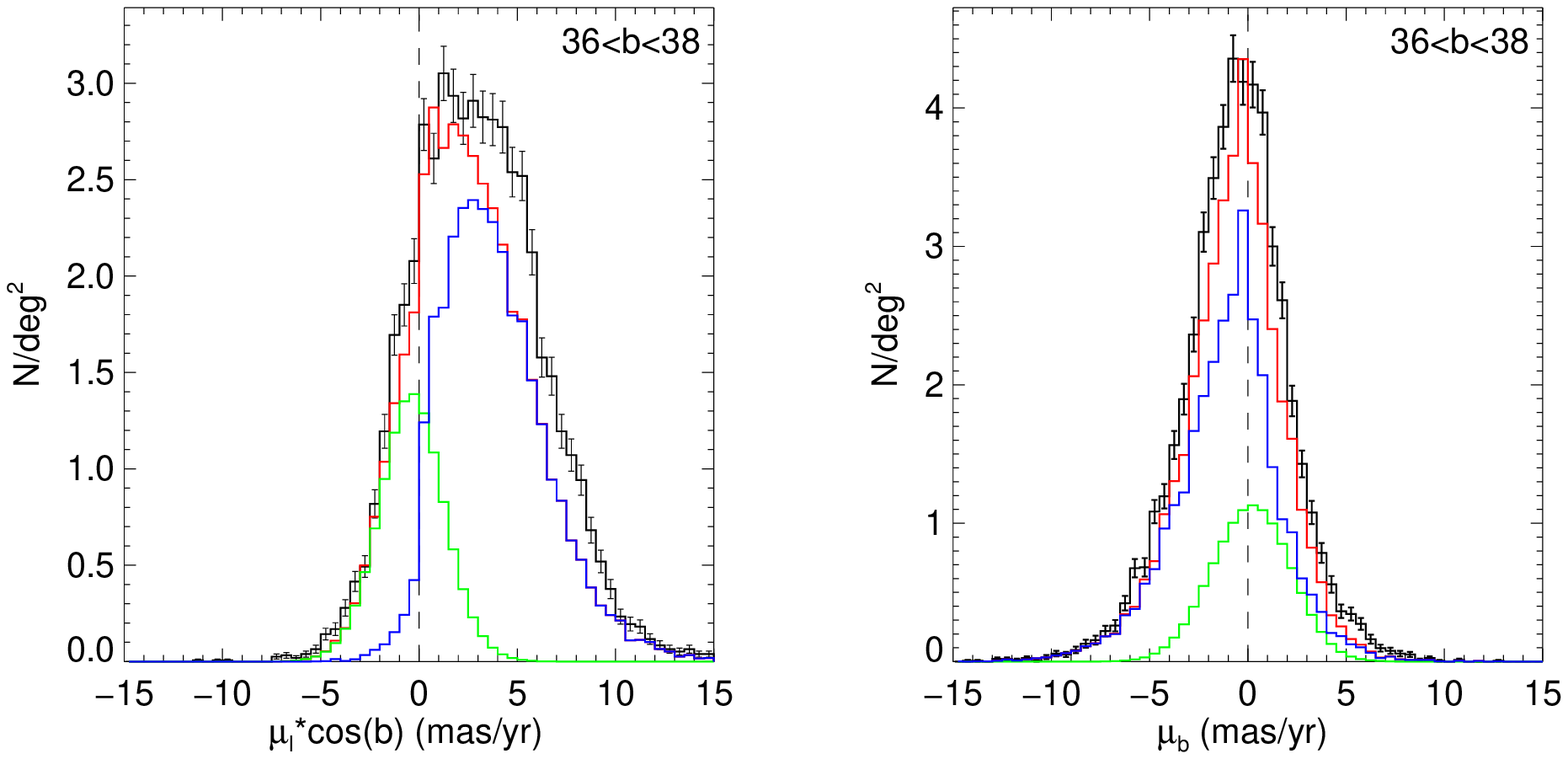}
\caption{Overview of Monoceros model fitting results, for fixed latitude data slices covering a Galactic longitude between $150<l<240$ deg. For each value of latitude, the left panel shows histograms in $\mu_{l}$ and right panels show $\mu_{b}$. Black lines indicate the histogram of observed data along with Poisson error bars. The red histograms show the best-fit models, composed of the best-fit toy model (in green) for Monoceros and the MW contamination from Galaxia (blue histogram). \label{Mono_resids}}
\end{figure*}

To test our cylinder models, we apply them to a sample of thin disk stars extracted from a low latitude SDSS stripe between $200<l<205$ deg and $5<b<10$ deg. We select a clean sample of thin disk stars using an isochrone mask with an age of 5 Gyr and [Fe/H]=$-$0.5 dex, adopting a distance of 3 kpc. We fit for the peak and standard deviation in v$_{\phi}$, v$_{z}$ space using least squares minimisation of the proper motion histograms. The observed distribution (in black) and resulting best-fit model (in red) are shown in Figure~\ref{test_thindisk}. The model is an excellent fit to the observed proper motion distribution, and the best-fit parameters (v$_{\phi}$=236.6$\pm$0.1 km/s, $\sigma_{v_\phi}$=13.9$\pm$0.1 km/s, v$_{z}$=-3.4$\pm$0.2 km/s, $\sigma_{v_z}$=11.5$\pm$0.2 km/s) are a good match to the parameters we expect for thin disk stars. Therefore, we conclude that the models are able to correctly reproduce the velocities of stars moving on a circular orbit around the MW.

\section{Velocities of the Monoceros features}
\label{Monofitting}
To obtain the velocity profiles of the Monoceros stream as a function of Galactic latitude, we first split the observed data into bins of 2 deg wide covering the range of $16$ $-$ $40$ deg in latitude~(see also Figure~\ref{Mono_densitypm}). For each slice, we use the model grid described in Section~\ref{Monomodels} to generate models as binned two dimensional Gaussians in v$_{\phi}$, v$_{z}$ space and fit for the peak and standard deviation parameters using least squares minimisation. When fitting the models to the Monoceros data, we also allow for contamination due to MW thick disk and halo stars. Therefore, we also fit for a MW contamination fraction using proper motion distributions determined from the Galaxia models~(see also Figure~\ref{Mono_pmhists}). In summary, we fit for the following parameters: v$_{\phi}$, $\sigma_{v_\phi}$, v$_{z}$, $\sigma_{v_z}$, f$_{MW}$.

Figure~\ref{Mono_resids} shows the observed proper motion distribution for each latitude slice~(black histogram), along with the best-fit model~(red histogram). The figure shows that overall the models provide a good fit to the data, with the peak and width of proper motion distributions correctly reproduced. Some disagreement between model and data is visible at the high proper motion tails of the $\mu_{l}$ distributions, with the mismatch increasing in severity toward higher latitudes. This effect is likely linked to residual MW contamination, which dominates at high positive proper motions. Figure~\ref{Mono_resids} shows that the asymmetry of the $\mu_{b}$ distributions is better reproduced by the models, with smaller mismatches throughout.

\begin{figure*}
\centering
\includegraphics[angle=0, width=0.95\textwidth]{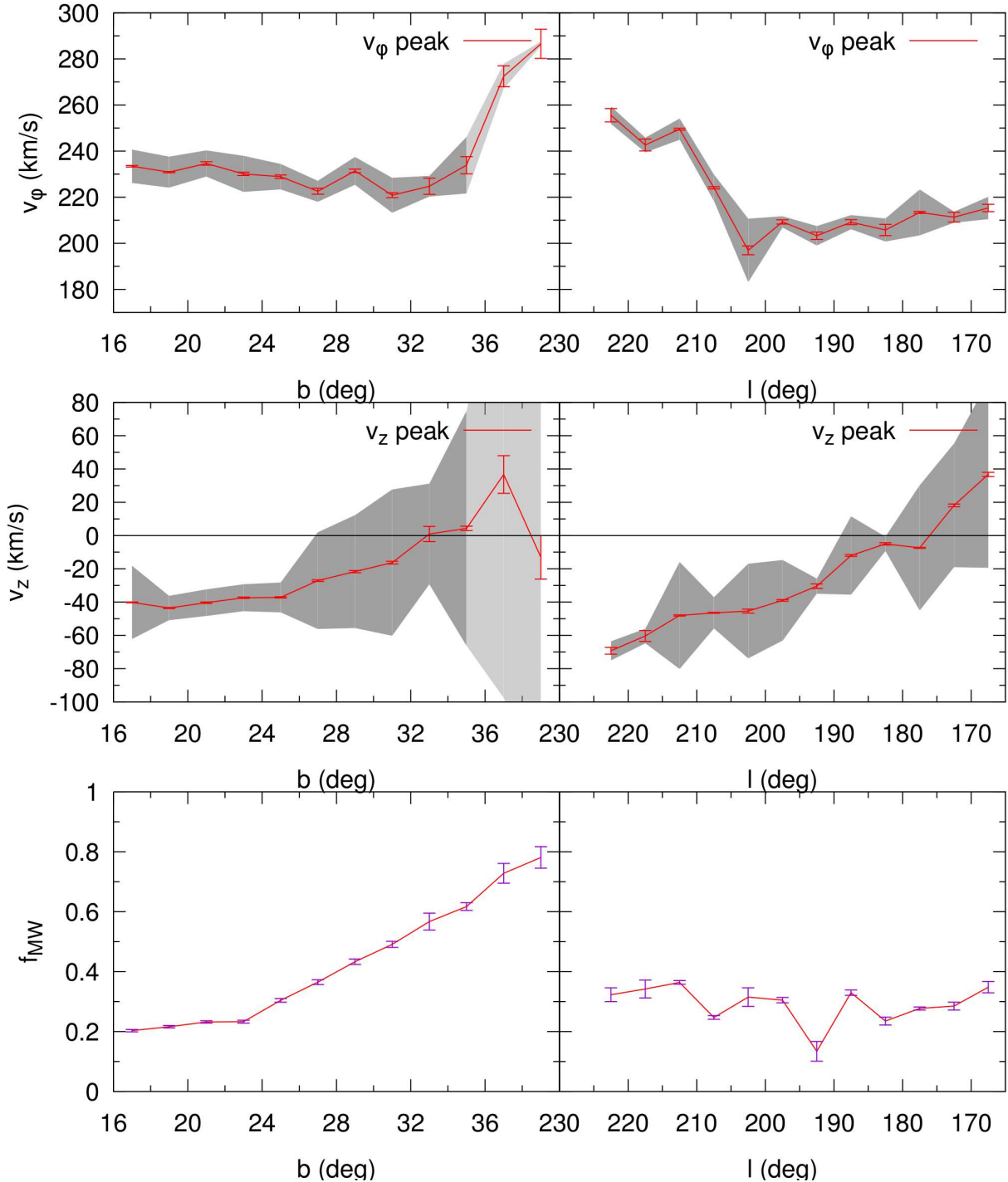}
\caption{Best-fitting parameters of the Monoceros models, as a function of Galactic latitude between 150$<$l$<$240 deg (left panels) and as function of Galactic longitude between 16$<$b$<$35 deg (right panels). The upper panels show results for the tangential velocity~(v$_{\phi}$) while the middle panels shows velocity in the z direction. Red lines indicate the central velocity of the best-fit models together with its uncertainty, while the grey shaded region indicates the velocity dispersion. Finally, the bottom panels show the fraction of MW contamination as a function of latitude. For latitudes with low Monoceros signal, shaded regions have been made transparent. \label{Mono_fitpars}}
\end{figure*}

The parameters of the best-fit models for each slice are shown as a function of latitude in the left panels of Figure~\ref{Mono_fitpars}, and listed in Table~\ref{Mono_parstable}. As evidenced by Figure~\ref{Mono_resids} the fraction of MW contamination in the data increases strongly as a function of latitude, increasing from a low $\approx$20 percent to nearly $80$ percent at high Galactic latitudes. This is consistent with Figure~\ref{Mono_densitypm} which shows a steadily decreasing fraction of the sky covered by prominent Monoceros features above $b=25$\,deg, and an almost complete lack of Monoceros above $b=35$\,deg. 

\begin{table*}
\caption[]{Parameters for each Galactic latitude and longitude slice for the best-fit Monoceros models shown in Figure~\ref{Mono_fitpars}. Peak values indicate the centre velocity of the best-fit model, and $\sigma$ indicates the velocity dispersion. Finally,  f$_{MW}$ shows the MW contamination fraction determined from the joint fit of the Monoceros model and Galaxia distribution. \label{Mono_parstable}}
\begin{center}
%\resizebox{0.45\textwidth}{!}{
\begin{tabular}{cccccc}
\hline\hline
b &  v$_{\phi}$ peak & 
$\sigma_{v_\phi}$ & v$_{z}$ peak & $\sigma_{v_z}$ & f$_{MW}$ \\
deg & km/s & km/s & km/s & km/s & \\
\hline
 16$-$18 & 233.43$\pm$0.32 &  7.24$\pm$2.75 & -40.17$\pm$0.25 &  22.01$\pm$0.73 & 0.20$\pm$0.01 \\
 18$-$20 & 230.87$\pm$0.26 &  6.74$\pm$1.05 & -43.54$\pm$0.30 &   7.33$\pm$1.38 & 0.22$\pm$0.01 \\
 20$-$22 & 234.62$\pm$0.75 &  5.70$\pm$2.42 & -40.34$\pm$0.45 &   8.01$\pm$2.26 & 0.23$\pm$0.01 \\
 22$-$24 & 230.11$\pm$0.71 &  7.83$\pm$3.05 & -37.39$\pm$0.42 &   8.11$\pm$3.04 & 0.23$\pm$0.01 \\
 24$-$26 & 228.90$\pm$0.77 &  5.52$\pm$1.90 & -37.17$\pm$0.38 &   8.97$\pm$3.35 & 0.30$\pm$0.01 \\
 26$-$28 & 222.58$\pm$1.29 &  4.56$\pm$1.97 & -27.12$\pm$0.50 &  29.07$\pm$1.18 & 0.37$\pm$0.01 \\
 28$-$30 & 231.41$\pm$0.76 &  6.04$\pm$3.70 & -21.67$\pm$0.66 &  33.91$\pm$1.45 & 0.43$\pm$0.01 \\
 30$-$32 & 220.85$\pm$1.06 &  7.58$\pm$6.63 & -16.29$\pm$0.84 &  43.98$\pm$1.52 & 0.49$\pm$0.01 \\
 32$-$34 & 224.74$\pm$3.50 &  4.49$\pm$5.13 &   0.92$\pm$4.58 &  30.25$\pm$25.88 & 0.57$\pm$0.03 \\
 34$-$36 & 233.87$\pm$3.72 & 12.26$\pm$12.66 &  4.30$\pm$1.32 &  70.47$\pm$2.08 & 0.62$\pm$0.01 \\
 36$-$38 & 272.47$\pm$4.53 &  5.39$\pm$6.58 &  36.68$\pm$11.32 & 134.20$\pm$23.15 & 0.73$\pm$0.03 \\
 38$-$40 & 286.55$\pm$6.31 &  1.00$\pm$9.26 & -13.09$\pm$13.07 & 150.00$\pm$0.00 & 0.78$\pm$0.04 \\
\hline\hline
l &  v$_{\phi}$ peak & $\sigma_{v_\phi}$ & v$_{z}$ peak & $\sigma_{v_z}$ & f$_{MW}$ \\
deg & km/s & km/s & km/s & km/s & \\
\hline
165$-$170 & 215.32$\pm$1.61 &  4.84$\pm$6.79 &  36.72$\pm$1.30 & 56.12$\pm$2.15 & 0.35$\pm$0.02 \\
170$-$175 & 211.31$\pm$2.11 &  2.39$\pm$2.14 &  18.12$\pm$0.80 & 37.16$\pm$1.63 & 0.29$\pm$0.01 \\
175$-$180 & 213.38$\pm$0.43 &  9.96$\pm$3.90 &  -7.38$\pm$0.31 & 37.60$\pm$0.57 & 0.28$\pm$0.01 \\
180$-$185 & 205.77$\pm$2.44 &  5.02$\pm$1.13 &  -5.04$\pm$0.71 &  4.39$\pm$2.09 & 0.24$\pm$0.01 \\
185$-$190 & 209.15$\pm$1.08 &  3.08$\pm$2.27 & -12.01$\pm$0.53 & 23.52$\pm$1.51 & 0.33$\pm$0.01 \\
190$-$195 & 203.26$\pm$1.59 &  4.21$\pm$3.33 & -30.41$\pm$1.38 &  4.55$\pm$2.92 & 0.13$\pm$0.03 \\
195$-$200 & 209.27$\pm$0.92 &  2.45$\pm$1.64 & -38.97$\pm$0.51 & 24.23$\pm$1.42 & 0.31$\pm$0.01 \\
200$-$205 & 196.93$\pm$1.93 & 13.73$\pm$6.67 & -45.41$\pm$1.20 & 28.42$\pm$2.81 & 0.32$\pm$0.03 \\
205$-$210 & 224.07$\pm$0.48 &  5.75$\pm$2.04 & -46.46$\pm$0.28 &  9.41$\pm$3.15 & 0.25$\pm$0.01 \\
210$-$215 & 249.52$\pm$0.41 &  4.60$\pm$2.81 & -48.06$\pm$0.39 & 32.14$\pm$0.81 & 0.36$\pm$0.01 \\
215$-$220 & 242.64$\pm$2.58 &  3.21$\pm$1.83 & -60.38$\pm$3.28 &  4.42$\pm$1.91 & 0.34$\pm$0.03 \\
220$-$225 & 255.56$\pm$2.89 &  3.80$\pm$2.43 & -69.25$\pm$2.04 &  5.72$\pm$2.28 & 0.32$\pm$0.02 \\
\hline 
\end{tabular}
%}
\end{center}
\end{table*}

When considering tangential velocities, Figure~\ref{Mono_fitpars} shows little variation for latitudes below $b=25$\,deg, followed by a slow decline towards higher latitudes. The dispersion of the Gaussian distribution in v$_{\phi}$ is generally small, as indicated by the narrowness of the $\mu_{l}$ peaks in Figure~\ref{Mono_pmhists}. The peak tangential velocity of the Monoceros Ring is $\approx 230 $\,km/s, which is almost 10 km/s lower than the assumed tangential velocity at the solar radius~(v$_{\phi,\odot}=239.5$\, km/s~\citet{Reid04}). The rotation curve of the MW is generally expected to drop at large Galactocentric radii, although the behaviour at intermediate radii is more uncertain~\citep{Kalberla07}. In particular, beyond the solar radius out to $20$\,kpc the rotation curve is expected to be influenced by the flaring of the observed \ion{H}{i} disk, which could lead to a rise of v$_{\phi}$ at the distances studied here~\citep{Wouterloot90,Merrifield92}. The tangential velocities of Monoceros are well within the range of possible velocities for MW disk rotation. Above $b=35$\,deg, tangential velocities vary greatly, which is likely caused by the strong MW halo contamination. In these cases, the model will attempt to fit the halo dominated signal using a cylindrical shell at a distance of $10$\,kpc, resulting in an unconstrained tangential velocity and very large width for any converged fits, as indeed we see in Figure~\ref{Mono_fitpars} at high latitudes. 

The vertical velocity of the Monoceros models shows a strong change as function of Galactic latitude, changing from $\approx-40$\,km/s at low latitudes and velocities comparable to zero at high latitudes as Monoceros reaches its highest point above the MW mid plane. The strong, but gradual change in v$_{z}$ indicates the velocity distribution of Monoceros smoothly changes as a function of latitude, without strong jumps or breaks. Given the clear change in density in Figure~\ref{Mono_densitypm} and change of dominance from Monoceros ACS to Ring component, we conclude that different Monoceros features are comprised of a similar velocity distribution.

Next we study the behaviour of velocities as a function of Galactic longitude. We split the Monoceros sample into bins of 5 deg wide covering the range of $165-225$ deg where Monoceros is clearly present (see Figure~\ref{Mono_densitypm}). Once again, for each slice we fit models in v$_{\phi}$, v$_{z}$ space using a least squares minimisation allowing for the possibility of contamination due to MW thick disk and halo stars. The right panels of Figure~\ref{Mono_fitpars} show the best-fit velocity parameters as a function of longitude, as well as the MW contamination fraction. The MW contamination fraction is always below $0.5$, consistent with the strong presence of Monoceros at all considered longitudes. \\
The behaviour of tangential velocity appears to separate into a low velocity component ($\approx 200$\,km/s) at low longitudes and a high velocity component ($\approx 250$\,km/s) at high longitudes, transitioning at $l=210$\,deg. Comparison to the bottom left panel of Figure~\ref{Mono_densitypm} shows that longitudes above $210$ deg mostly sample low latitude regions dominated by the ACS component, while longitudes below 210 deg sample both components over a wide range of latitudes. This could be an indication that the two Monoceros components rotate with different tangential velocities. However, at lowest longitudes ($<$175 deg) the signal is dominated by the ACS component, while the tangential velocities are consistent with those of the intermediate longitude range. Therefore, it is not clear how the velocities of both components are related.

\begin{figure*}
\centering
\includegraphics[angle=0, width=0.95\textwidth]{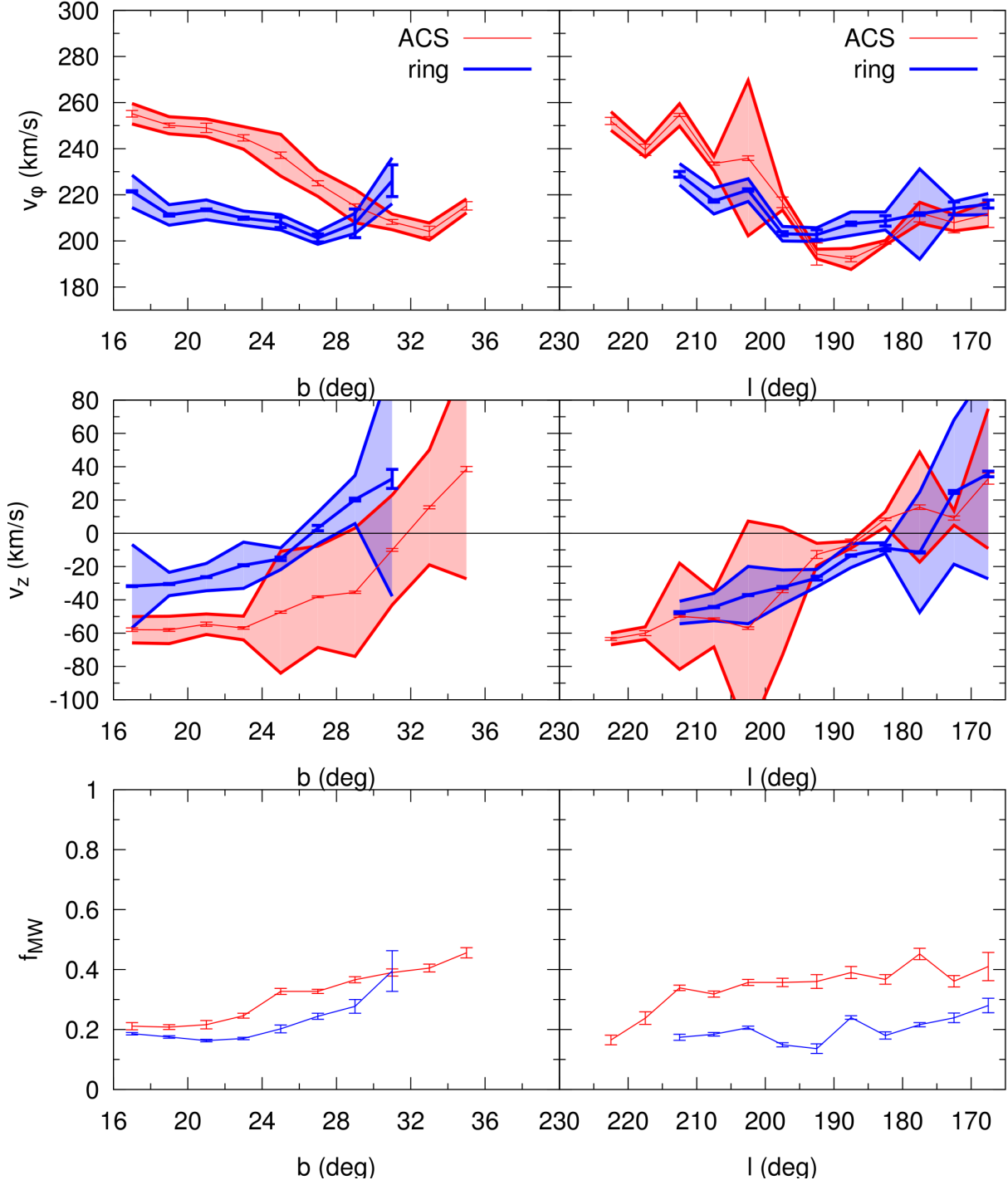}
\caption{Best-fitting parameters for the lower latitude Ring component and the higher latitude stream component of the Monoceros complex (shown in blue and red respectively), as a function of Galactic latitude and longitude. See Figure~\ref{Mono_fitpars} for a detailed description of different panels.\label{Mono_fitpars_components}}
\end{figure*}

When considering the vertical velocity, we once again see a clear trend from strongly negative velocity to mildly positive velocity, this time with Galactic longitude. The range of vertical velocity is larger, ranging from $-70$\,km/s at the high longitude end to $+40$\,km/s at the low longitude end. This clearly indicates that the vertical velocity is dependent on both longitude and latitude. No clear change in the trend is visible, despite the fact that different longitude bins are dominated by different Monoceros features.

\subsection{Separating the ACS and Ring components}
To investigate the velocities of the different Monoceros components in more detail, we separate the Ring and ACS features and investigate the velocities as a function of latitude separately. We select the ACS sample by making use of the great circle fits of \citet{Grillmair06}, which are offset in each direction by 5 deg, resulting in the two blue great circles displayed in the bottom left panel of Figure~\ref{Mono_densitypm}. The ACS sample is selected as the region enclosed by the two great circles, while the Ring sample is composed of all stars below the bottom blue great circle. Subsequently, each sample is divided into slices of constant latitude and longitude, for which the Monoceros models are fit in the same way as for the full sample. Figure~\ref{Mono_fitpars_components} shows the best-fitting peak and width parameters for tangential and vertical velocity for each component, along with the MW background fraction.

Similar to Figure~\ref{Mono_fitpars}, we once again see a clear sequence of slowly declining tangential velocity with latitude. However, the two components of Monoceros show similar sequences offset from each other by $\approx 40$\,km/s with the ACS rotating faster. The two Anti-centre components also separate from each other in vertical velocity v$_z$ as shown in the middle panels of Figure~\ref{Mono_fitpars_components}. However, the right panels of Figure~\ref{Mono_fitpars_components} show that both components display similar velocities as function of longitude, separating only at the highest longitudes where only the ACS component is strongly present. Therefore, we conclude that a more global pattern of vertical and tangential velocity is present across the Monoceros complex.

As expected, the sequence in Figure~\ref{Mono_fitpars} corresponds to the average of the two components shown in Figure~\ref{Mono_fitpars_components}. The difference between the two components also explains the behaviour seen in Figure~\ref{Mono_fitpars} as a function of longitude. At the highest longitudes, only the ACS stream is present, which rotates at $\approx 250$\, km/s at the low latitudes sampled. At intermediate longitudes we sample both components, resulting in a much lower average velocity of $\approx 200$\,km/s.

\begin{figure}
\centering
\includegraphics[angle=0, width=0.495\textwidth]{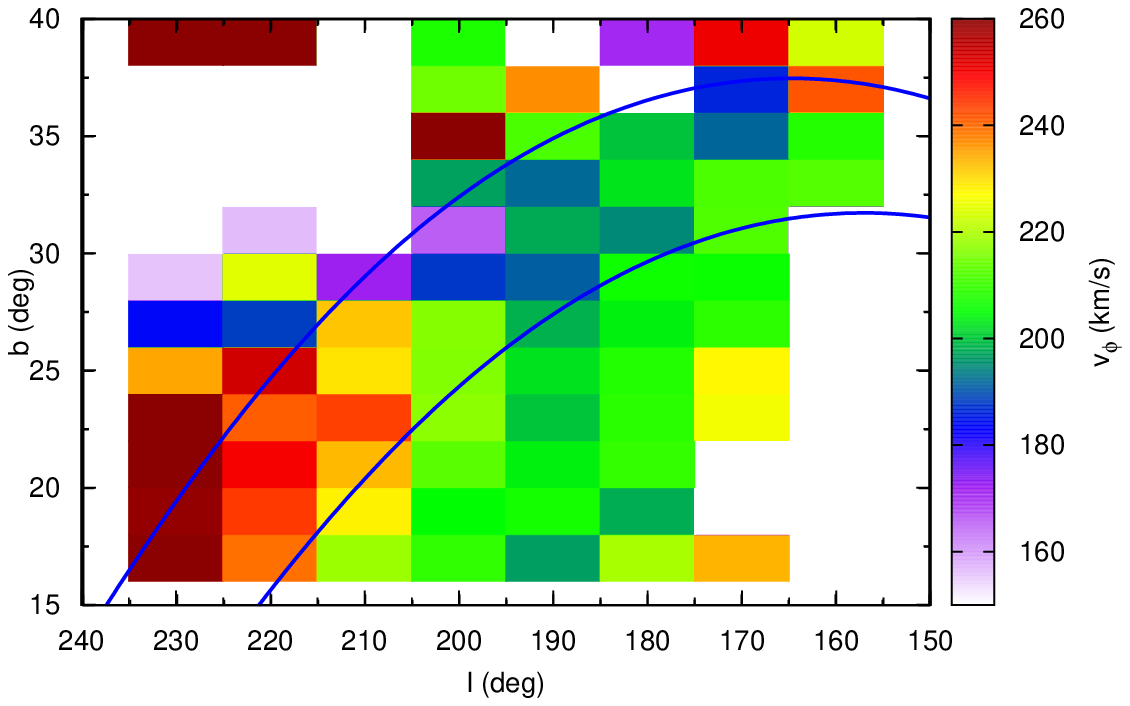}
\includegraphics[angle=0, width=0.495\textwidth]{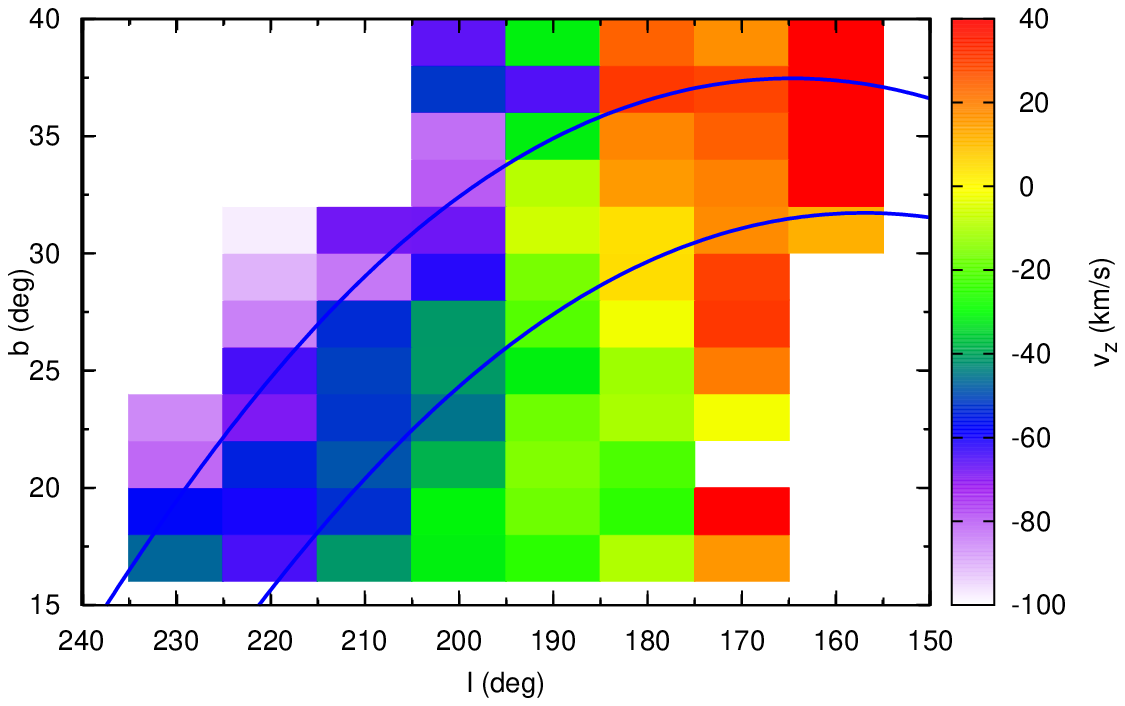}
\includegraphics[angle=0, width=0.495\textwidth]{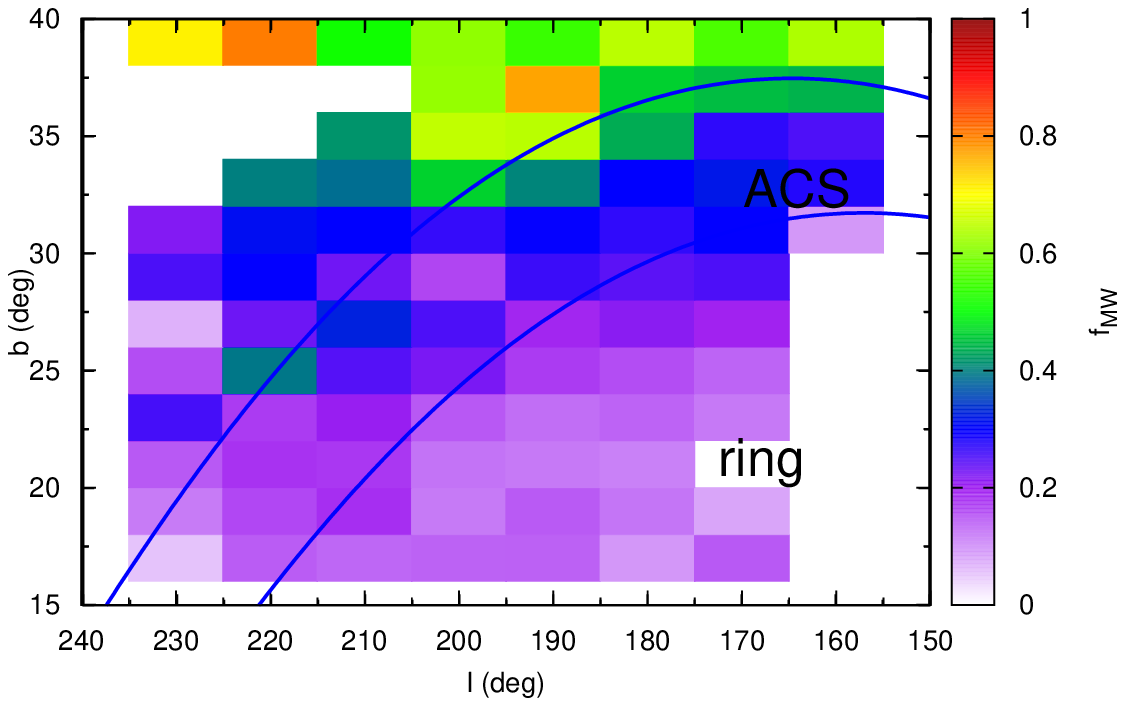}
\caption{Best-fitting parameters of our Monoceros models as a function of Galactic longitude and latitude using 10$\times$2 deg bins. Similar to Figure~\ref{Mono_fitpars}, the upper panel shows the tangential velocity~(v$_{\phi}$), the middle panel the vertical velocity and the bottom panel the fraction of MW contamination. The blue lines indicate the area of sky associated with the ACS stream, while the area below the blue lines is occupied by the Monoceros Ring. \label{Mono_spatialfits}}
\end{figure}

\section{Spatial fits of Monoceros velocities}
\label{spatfits}
We now study the velocity distribution of Monoceros as a function of both Galactic longitude and latitude, to decouple variations with longitude from the spatial location of individual Monoceros sequences. In this way, we can investigate variations within both the ACS and Ring structures as a function of longitude, and avoid biases at low latitudes due to the high density of the Monoceros Ring features at lower longitudes. We divide our sample into bins of 10$\times$2 deg and fit Monoceros models to each pixel separately. Once again, we also fit for a contamination fraction using synthetic MW stars generated using the Galaxia models. \\
Figure~\ref{Mono_spatialfits} shows the best-fitting parameters for each spatial bin, with tangential velocity~(v$_{\phi}$) in the upper panel, vertical velocity in the middle panel and MW disk/halo contamination fraction in the bottom panel. The contamination fraction clearly changes as a function of longitude and latitude predominantly following the density change of Monoceros stars. In areas of high contamination~(and therefore very low or negligible Monoceros density), the Monoceros models fit instead the MW background using a cylindrical shell rotating model, resulting in parameters values limited to the edges of our allowed grid. \\

Similar to Figure~\ref{Mono_fitpars}, vertical velocity in Figure~\ref{Mono_spatialfits} shows a clear dependence on latitude, with stronger negative velocities at low latitudes. However, Figure~\ref{Mono_spatialfits} also reveals a dependence on Galactic longitude for both vertical and tangential velocities. If we separate parts of the sky associated to the ACS stream~(as shown by the blue lines in Figure~\ref{Mono_spatialfits}) and the canonical Monoceros Ring, we can investigate trends for both components. Indeed, Figure~\ref{Mono_spatialfits} shows that the tangential velocity of the ACS component is higher than for the Ring, both at high and low latitudes. In general, as we move from the low longitude, low latitude Ring dominated region to the ACS dominated region the tangential velocity gradually increases. \\
In terms of vertical velocity, a clear sequence is apparent in the ACS, with the low longitude parts showing positive velocities~(moving up from the MW mid plane) followed by a phase of zero vertical velocity as the stream reaches its highest point and consecutively more negative velocity as the longitude increases and latitude decreases (falling back down to the mid plane). The behaviour of the Ring component is less clear, with mostly negative vertical velocities. However, the same overall trend seems to be present but less pronounced. There is no clear sign of different velocity behaviour between the two components in the vertical direction. Instead, both are consistent with disk-like motion, at the same distance from the Sun and with similar metallicities.

\section{Conclusions and discussion}
\label{conclusions}

In this work, we have utilised re-calibrated stellar positions from
SDSS Data Release 10~\citep{Ahn14} and the {\it Gaia}
satellite~\citep{GAIAmain1,GAIAmain2} to study the proper motions of
the Monoceros Ring in the direction of the Galactic
anti-centre. Typically, for an individual star in the SDSS-{\it Gaia}
catalog, we estimate the random proper motion error to be between 1
and 2 mas/year depending on the temporal baseline and the apparent
magnitude. The systematic error is demonstrated not to exceed 0.1
mas/year. By selecting stars belonging to the overdense MS at a
distance of $\approx$10 kpc~(see Figure~\ref{mono_CMD}) we have
obtained a sample of stars predominantly belonging to the Monoceros
feature. The spatial density distribution of our cleaned tracer sample
reproduces all of the known features of Monoceros, including the low
latitude Ring and the higher latitude ACS stream.

For simplicity, we have adopted a fixed distance for the different
Monoceros components in this work, and have not attempted to fit for
any distance variation across the anti-centre region. However, earlier
studies of SDSS photometric data have shown that the distance gradient
of the Monoceros Ring and ACS is small~\citet{Grillmair08, Li12}, and
our colour-magnitude selection is sufficiently wide to always include
the dominant MS feature across the region studied~(see
Figure~\ref{mono_CMD}). Furthermore, small~($\approx$10\% or 1kpc)
differences in the adopted distance will not result in greatly
different proper motion peak values. Henceforth, we conclude that
taking into account small distance variations will not greatly alter
the conclusions presented in this work.

Our selected sample shows a clear proper motion signature of
Monoceros, which correlates perfectly with features in the spatial
density maps (as shown in Figure~\ref{Mono_densitypm}). Monoceros
stars produce an obvious peak in proper motions at low Galactic
latitude, where this structure is expected to dominate the
sample. Comparison to the Galaxia model~\citep{Sharma11} shows that
the position of this peak is not consistent with the proper motion
distribution of the conventional MW components at this
location (see Figure~\ref{Mono_pmhists}).

The centroid of the proper motion distribution of the Monoceros stars
exhibits a clear change as a function of Galactic latitude~(see
Figure~\ref{Mono_pmhists}), indicating a smooth change in the velocity
of Monoceros with height above the Galactic plane. This is the first
time that such a subtle proper motion signal of Monoceros has been
uncovered. Figure~\ref{Lidist_comp} shows that it is only due to the
greater accuracy of the SDSS-{\it Gaia} catalog that we are able to
robustly distinguish and characterise multiple components in the
proper motion distribution. It appears that in previously available
wide-area proper motion datasets, the low-level evolution of the peak
values with latitude~($\approx$2\,mas/yr in $\mu_{l}$ and
$\approx$0.5\,mas/yr in $\mu_{b}$) is easily wiped out, thus
preventing the detection of the gradually changing Monoceros
properties.

%This also highlights the power of {\it Gaia} in obtaining homogeneous,
%accurate proper motions across a large area of the sky.

We approximate the proper motion distributions using a cylinder-like
model centred on the MW with a distance of $10$\,kpc from the Sun, in
which we generate stellar velocities as drawn from a two Gaussian
distributions in v$_{\phi}$ and v$_{z}$ space. The models generally
produce a good fit to the observed proper motion histograms, apart
from small systematic residuals at large positive $\mu_{l}$, where the
distribution is asymmetric and dominated by the MW contribution. These
mismatches are likely due to our fairly simple model, which is neither
tilted with respect to the Galactic plane, nor takes into account
asymmetric drift and irregularities in the Monoceros
structures. Furthermore, it is also quite likely that the distribution
of the MW halo stars is different from that posited in
Galaxia~\citep[see e.g.][]{Deason11,xue2015}.

Our best model implies that Monoceros is spinning with a tangential
velocity of $\sim 230$ kms$^{-1}$, only slightly lower than of the
Sun. Therefore, in line with a number of previous studies, we have
shown conclusively that Monoceros is moving on a nearly circular
prograde orbit, which rules out tidal debris models that rely on a
retrograde satellite in-fall, such as those of~\citet{Sheffield14} and
the retrograde model of~\citet{Penarrubia05}. Measurements of the
rotation curve of the MW at Galactocentric radii covered by our
Monoceros sample cover a range of possible rotation velocities for the
MW disk are possible~\citep{Kalberla07}. Beyond the solar radius, out
to $20$\,kpc the rotation curve is expected to be influenced by the
flaring of the observed \ion{H}{i} disk, which can lead to a rise of
v$_{\phi}$ at the distances studied
here~\citep{Wouterloot90,Merrifield92}. The tangential velocities of
Monoceros we determine in this work are well within the range of
possible velocities for MW disk rotation.

The vertical velocities of our Ring models (see
Figures~\ref{Mono_fitpars} and~\ref{Mono_spatialfits}) also reveal
that stars in the Monoceros structure are moving up from the MW mid
plane at the lowest longitudes studied here ($150<l<170$), before
falling back down toward the Galactic plane at higher longitudes
($l>180$), with a vertical velocity that gradually becomes more
negative toward lower latitudes. This systematic vertical motion as
function of longitude is reminiscent of a wave or a ripple propagating
through the MW disk. Figure~\ref{Mono_fitpars_components} shows that
the amplitude of the tangential velocity gradients is greater for the
higher latitude so-called ACS component than for the lower latitude
Ring. However, both structural components display similar velocity
trends overall.  Note that according to Figure~\ref{Mono_spatialfits}
there are changes in the behaviour of the velocity distribution at the
boundary between the ACS and the Ring components, but these appear
relatively minor. Given the broad-brush similarities in kinematics, it
is likely that the ACS and the Ring formation mechanisms are
inter-related.

Given the results presented in this work, we can try to shed some new
light on the origin of the Monoceros feature. In particular, we can
try to distinguish between the two main scenarios proposed to explain
the Monoceros Ring: dwarf galaxy debris and disk disturbance. In the
case of a dwarf debris origin, different Monoceros sub-structures
would represent (superpositions of) multiple wraps of the stellar
stream generated by the disruption of the progenitor galaxy~\citep[see
  e.g.][]{Penarrubia05}. In this model, in bins of Galactic longitude
each wrap of the stream would produce a distinct peak in $v_z$, thus
creating a multi-modal distribution. Is this supported by the data
presented here?  Certainly, as evidenced in
Figures~\ref{Mono_fitpars_components} and ~\ref{Mono_spatialfits},
there exist values of $l$ where at least two different $v_z$ peaks are
reported. However, observations appear to show great coherence in
terms of velocity patterns as a function of position, with multiple
peaks - even if detected - not too far from each other in the velocity
space. Superficially, in favour of the dwarf debris scenario are also
the low velocity dispersions measured here. These however must be
taken with a pinch of salt as we do not possess actual proper motion
errors for individual stars, and use instead a rough approximation
based on the QSO astrometry. We envisage that this approach is robust
enough to study the evolution of the proper motion centroid, but might
be too crude to measure the width of the distribution. Another
relevant consideration is whether the Monoceros properties inferred in
our work are compatible with the only currently known progenitor of
Monoceros, i.e. the Canis Major overdensity~\citep{Martin04}. The
orbit of Canis Major is indeed very circular~(e=0.16, according
to~\citet{Momany04}) and it is located just below the MW plane at a
distance of $\approx$7\,kpc. However, the tangential velocity of Canis
Major inferred from radial velocities is below that of the
Sun~\citep{Martin04}, somewhere between 160 and 200 kms$^{-1}$, unlike
the tangential velocity of Monoceros derived here. Similarly, the
rotation velocity of Tri-And is estimated to be around $\sim160$
kms$^{-1}$ \citep[see][]{deason_triand}, much lower compared to our
measurements for Monoceros.

In the case of a disk origin, the Monoceros Ring is the result of a
disturbance in the MW disk induced by the repeated fly-bys of a
massive dark matter
subhalo~\citep[e.g.][]{Kazantzidis08,Gomez16}. Stars in the outer disk
are kicked up to high Galactic latitudes, forming structures with a
morphology dependent on the properties and orbit of the perturbing
system. In this scenario, Monoceros Ring could possibly be related to
the infall of the massive Sagittarius dwarf
galaxy~\citep{Jiang00,Purcell11,Gibbons17}. It appears possible for
the disturbance to be confined to a relatively small range of
distances and form shell-like features reminiscent of what is actually
observed in the Anti-centre~\citep{helmi2003,Kazantzidis08,
  michel2011}. Within this scenario, the vertical velocity gradient
measured here can be interpreted as a signature of a wave of stars, a
ripple propagating through the disc. Moreover, the inferred tangential
velocities of Monoceros are easily achieved for stars already moving
with the MW disk rotational velocity. The metallicities of the
Monoceros stars are well constrained from spectroscopic observations
and are generally found to be ${\rm [Fe/H]} \approx-0.8$\,dex,
i.e. more metal-poor than the thin disk. However, this does not rule
out a thin-disk origin for Monoceros, given the presence of a radial
metallicity gradient in the thin disk, which can easily lead to a
$\approx 0.5 $\,dex metallicity difference at a distance of
$10$\,kpc~\citep{Cheng12}. There are, however, complications with this
scenario just as well. Most importantly, it is the fact that the
perturber, say the Sagittarius dwarf, must have gone through the MW
disc a number of times, thus giving rise not just to one but to many
waves. It is not clear whether the kinematic pattern resulting from
the superposition of these waves should appear as coherent as observed
here.

Even when tested with detailed tangential kinematics, the Monoceros
Ring appears to have properties that require contribution from both
formation scenarios. Consequently, could this structure be a
superposition of the kicked disc stars and the tidal debris from a
dwarf galaxy on low-inclined orbit? It should be possible to answer
this question with the help of the upcoming astrometric data from the
{\it Gaia} satellite complemented by the all-sky spectroscopy from
WEAVE, DESI and 4MOST.

\section*{Acknowledgements}
The authors have enjoyed many a conversation with the members of the
Cambridge Streams Club. We also thank Jorge Pe{\~n}arrubia and Chervin
Laporte for valuable discussions.

The research leading to these results has received funding from the
European Research Council under the European Union's Seventh Framework
Programme (FP/2007-2013) / ERC Grant Agreement
n. 308024. T.d.B. acknowledges financial support from the ERC. SK
thanks the United Kingdom Science and Technology Council (STFC) for
the award of the Ernest Rutherford fellowship (grant number
ST/N004493/1).

Funding for SDSS-III has been provided by the Alfred P. Sloan
Foundation, the Participating Institutions, the National Science
Foundation, and the U.S. Department of Energy Office of Science. The
SDSS-III web site is http://www.sdss3.org/.

SDSS-III is managed by the Astrophysical Research Consortium for the
Participating Institutions of the SDSS-III Collaboration including the
University of Arizona, the Brazilian Participation Group, Brookhaven
National Laboratory, Carnegie Mellon University, University of
Florida, the French Participation Group, the German Participation
Group, Harvard University, the Instituto de Astrofisica de Canarias,
the Michigan State/Notre Dame/JINA Participation Group, Johns Hopkins
University, Lawrence Berkeley National Laboratory, Max Planck
Institute for Astrophysics, Max Planck Institute for Extraterrestrial
Physics, New Mexico State University, New York University, Ohio State
University, Pennsylvania State University, University of Portsmouth,
Princeton University, the Spanish Participation Group, University of
Tokyo, University of Utah, Vanderbilt University, University of
Virginia, University of Washington, and Yale University.

This work has made use of data from the European Space Agency (ESA)
mission {\it Gaia} {\scriptsize{(\url{http://www.cosmos.esa.int /gaia})}}, \\
processed by the {\it Gaia} Data Processing and Analysis Consortium (DPAC,
{\scriptsize{\url{http://www.cosmos.esa.int/web/gaia/dpac/consortium}}}). Funding for the DPAC has been provided by national institutions, in particular
the institutions participating in the {\it Gaia} Multilateral Agreement.
\bibliographystyle{mn2e_fixed}
\bibliography{Bibliography}

\begin{thebibliography}{77}
\expandafter\ifx\csname natexlab\endcsname\relax\def\natexlab#1{#1}\fi

\bibitem[{{Abadi} {et~al}\mbox{.}(2003){Abadi}, {Navarro}, {Steinmetz}, \&
  {Eke}}]{abadi2003}
{Abadi} M.~G., {Navarro} J.~F., {Steinmetz} M., {Eke} V.~R., 2003, \apj, 597,
  21

\bibitem[{{Ahn} {et~al}\mbox{.}(2014){Ahn}, {Alexandroff}, {Allende Prieto},
  {Anders}, {Anderson}, {Anderton}, {Andrews}, {Aubourg}, {Bailey}, {Bastien},
  \& et~al.}]{Ahn14}
{Ahn} C.~P. {et~al.}, 2014, \apjs, 211, 17

\bibitem[{{Barnes} \& {Hernquist}(1992)}]{barnes1992}
{Barnes} J.~E., {Hernquist} L., 1992, \araa, 30, 705

\bibitem[{{Bellazzini} {et~al}\mbox{.}(2006){Bellazzini}, {Ibata}, {Martin},
  {Lewis}, {Conn}, \& {Irwin}}]{bella2006}
{Bellazzini} M., {Ibata} R., {Martin} N., {Lewis} G.~F., {Conn} B., {Irwin}
  M.~J., 2006, \mnras, 366, 865

\bibitem[{{Belokurov} {et~al}\mbox{.}(2007){Belokurov}, {Evans}, {Irwin},
  {Lynden-Bell}, {Yanny}, {Vidrih}, {Gilmore}, {Seabroke}, {Zucker},
  {Wilkinson}, {Hewett}, {Bramich}, {Fellhauer}, {Newberg}, {Wyse}, {Beers},
  {Bell}, {Barentine}, {Brinkmann}, {Cole}, {Pan}, \& {York}}]{Belokurov072}
{Belokurov} V. {et~al.}, 2007, \apj, 658, 337

\bibitem[{{Belokurov} {et~al}\mbox{.}(2014){Belokurov}, {Koposov}, {Evans},
  {Pe{\~n}arrubia}, {Irwin}, {Smith}, {Lewis}, {Gieles}, {Wilkinson},
  {Gilmore}, {Olszewski}, \& {Niederste-Ostholt}}]{Belokurov14}
{Belokurov} V. {et~al.}, 2014, \mnras, 437, 116

\bibitem[{{Benson} {et~al}\mbox{.}(2004){Benson}, {Lacey}, {Frenk}, {Baugh}, \&
  {Cole}}]{benson2004}
{Benson} A.~J., {Lacey} C.~G., {Frenk} C.~S., {Baugh} C.~M., {Cole} S., 2004,
  \mnras, 351, 1215

\bibitem[{{Bournaud}, {Jog} \& {Combes}(2007){Bournaud}, {Jog}, \&
  {Combes}}]{bournaud2007}
{Bournaud} F., {Jog} C.~J., {Combes} F., 2007, \aap, 476, 1179

\bibitem[{{Carlin} {et~al}\mbox{.}(2010){Carlin}, {Casetti-Dinescu},
  {Grillmair}, {Majewski}, \& {Girard}}]{Carlin10}
{Carlin} J.~L., {Casetti-Dinescu} D.~I., {Grillmair} C.~J., {Majewski} S.~R.,
  {Girard} T.~M., 2010, \apj, 725, 2290

\bibitem[{{Casey} {et~al}\mbox{.}(2013){Casey}, {Da Costa}, {Keller}, \&
  {Maunder}}]{Casey13}
{Casey} A.~R., {Da Costa} G., {Keller} S.~C., {Maunder} E., 2013, \apj, 764, 39

\bibitem[{{Cheng} {et~al}\mbox{.}(2012){Cheng}, {Rockosi}, {Morrison},
  {Sch{\"o}nrich}, {Lee}, {Beers}, {Bizyaev}, {Pan}, \& {Schneider}}]{Cheng12}
{Cheng} J.~Y. {et~al.}, 2012, \apj, 746, 149

\bibitem[{{Chou} {et~al}\mbox{.}(2010){Chou}, {Majewski}, {Cunha}, {Smith},
  {Patterson}, \& {Mart{\'{\i}}nez-Delgado}}]{chou2010}
{Chou} M.-Y., {Majewski} S.~R., {Cunha} K., {Smith} V.~V., {Patterson} R.~J.,
  {Mart{\'{\i}}nez-Delgado} D., 2010, \apjl, 720, L5

\bibitem[{{Chou} {et~al}\mbox{.}(2011){Chou}, {Majewski}, {Cunha}, {Smith},
  {Patterson}, \& {Mart{\'{\i}}nez-Delgado}}]{chou2011}
{Chou} M.-Y., {Majewski} S.~R., {Cunha} K., {Smith} V.~V., {Patterson} R.~J.,
  {Mart{\'{\i}}nez-Delgado} D., 2011, \apjl, 731, L30

\bibitem[{{Conn} {et~al}\mbox{.}(2005){Conn}, {Martin}, {Lewis}, {Ibata},
  {Bellazzini}, \& {Irwin}}]{conn2005}
{Conn} B.~C., {Martin} N.~F., {Lewis} G.~F., {Ibata} R.~A., {Bellazzini} M.,
  {Irwin} M.~J., 2005, \mnras, 364, L13

\bibitem[{{Crane} {et~al}\mbox{.}(2003){Crane}, {Majewski}, {Rocha-Pinto},
  {Frinchaboy}, {Skrutskie}, \& {Law}}]{Crane03}
{Crane} J.~D., {Majewski} S.~R., {Rocha-Pinto} H.~J., {Frinchaboy} P.~M.,
  {Skrutskie} M.~F., {Law} D.~R., 2003, \apjl, 594, L119

\bibitem[{{Deason}, {Belokurov} \& {Evans}(2011){Deason}, {Belokurov}, \&
  {Evans}}]{Deason11}
{Deason} A.~J., {Belokurov} V., {Evans} N.~W., 2011, \mnras, 416, 2903

\bibitem[{{Deason} {et~al}\mbox{.}(2014){Deason}, {Belokurov}, {Hamren},
  {Koposov}, {Gilbert}, {Beaton}, {Dorman}, {Guhathakurta}, {Majewski}, \&
  {Cunningham}}]{deason_triand}
{Deason} A.~J. {et~al.}, 2014, \mnras, 444, 3975

\bibitem[{{Deason} {et~al}\mbox{.}(2017){Deason}, {Belokurov}, {Koposov},
  {Gomez}, {Grand}, {Marinacci}, \& {Pakmor}}]{Deason17}
{Deason} A.~J., {Belokurov} V., {Koposov} S.~E., {Gomez} F.~A., {Grand} R.~J.,
  {Marinacci} F., {Pakmor} R., 2017, ArXiv e-prints

\bibitem[{{Eliche-Moral} {et~al}\mbox{.}(2011){Eliche-Moral},
  {Gonz{\'a}lez-Garc{\'{\i}}a}, {Balcells}, {Aguerri}, {Gallego}, {Zamorano},
  \& {Prieto}}]{eliche2011}
{Eliche-Moral} M.~C., {Gonz{\'a}lez-Garc{\'{\i}}a} A.~C., {Balcells} M.,
  {Aguerri} J.~A.~L., {Gallego} J., {Zamorano} J., {Prieto} M., 2011, \aap,
  533, A104

\bibitem[{{Gaia Collaboration} {et~al}\mbox{.}(2016{\natexlab{a}}){Gaia
  Collaboration}, {Brown}, {Vallenari}, {Prusti}, {de Bruijne}, {Mignard},
  {Drimmel}, {Babusiaux}, {Bailer-Jones}, {Bastian}, \& et~al.}]{GAIAmain1}
{Gaia Collaboration} {et~al.}, 2016{\natexlab{a}}, \aap, 595, A2

\bibitem[{{Gaia Collaboration} {et~al}\mbox{.}(2016{\natexlab{b}}){Gaia
  Collaboration}, {Prusti}, {de Bruijne}, {Brown}, {Vallenari}, {Babusiaux},
  {Bailer-Jones}, {Bastian}, {Biermann}, {Evans}, \& et~al.}]{GAIAmain2}
{Gaia Collaboration} {et~al.}, 2016{\natexlab{b}}, \aap, 595, A1

\bibitem[{{Gibbons}, {Belokurov} \& {Evans}(2017){Gibbons}, {Belokurov}, \&
  {Evans}}]{Gibbons17}
{Gibbons} S.~L.~J., {Belokurov} V., {Evans} N.~W., 2017, \mnras, 464, 794

\bibitem[{{Gillessen} {et~al}\mbox{.}(2009){Gillessen}, {Eisenhauer}, {Trippe},
  {Alexander}, {Genzel}, {Martins}, \& {Ott}}]{Gillessen09}
{Gillessen} S., {Eisenhauer} F., {Trippe} S., {Alexander} T., {Genzel} R.,
  {Martins} F., {Ott} T., 2009, \apj, 692, 1075

\bibitem[{{Gilmore} \& {Zeilik}(2000)}]{Gilmore00}
{Gilmore} G.~F., {Zeilik} M., 2000, {Star Populations and the Solar
  Neighborhood}, p. 471

\bibitem[{{G{\'o}mez} {et~al}\mbox{.}(2017){G{\'o}mez}, {Grand}, {Monachesi},
  {White}, {Bustamante}, {Marinacci}, {Pakmor}, {Simpson}, {Springel}, \&
  {Frenk}}]{gomez2017}
{G{\'o}mez} F.~A. {et~al.}, 2017, ArXiv e-prints

\bibitem[{{G{\'o}mez} {et~al}\mbox{.}(2016){G{\'o}mez}, {White}, {Marinacci},
  {Slater}, {Grand}, {Springel}, \& {Pakmor}}]{Gomez16}
{G{\'o}mez} F.~A., {White} S.~D.~M., {Marinacci} F., {Slater} C.~T., {Grand}
  R.~J.~J., {Springel} V., {Pakmor} R., 2016, \mnras, 456, 2779

\bibitem[{{Grillmair}(2006)}]{Grillmair06}
{Grillmair} C.~J., 2006, \apjl, 651, L29

\bibitem[{{Grillmair}(2011)}]{Grillmair11}
{Grillmair} C.~J., 2011, \apj, 738, 98

\bibitem[{{Grillmair}, {Carlin} \& {Majewski}(2008){Grillmair}, {Carlin}, \&
  {Majewski}}]{Grillmair08}
{Grillmair} C.~J., {Carlin} J.~L., {Majewski} S.~R., 2008, \apjl, 689, L117

\bibitem[{{Guo} \& {White}(2008)}]{guo2008}
{Guo} Q., {White} S.~D.~M., 2008, \mnras, 384, 2

\bibitem[{{Helmi} {et~al}\mbox{.}(2003){Helmi}, {Navarro}, {Meza}, {Steinmetz},
  \& {Eke}}]{helmi2003}
{Helmi} A., {Navarro} J.~F., {Meza} A., {Steinmetz} M., {Eke} V.~R., 2003,
  \apjl, 592, L25

\bibitem[{{Ibata} {et~al}\mbox{.}(2003){Ibata}, {Irwin}, {Lewis}, {Ferguson},
  \& {Tanvir}}]{Ibata03}
{Ibata} R.~A., {Irwin} M.~J., {Lewis} G.~F., {Ferguson} A.~M.~N., {Tanvir} N.,
  2003, \mnras, 340, L21

\bibitem[{{Ivezi{\'c}} {et~al}\mbox{.}(2008){Ivezi{\'c}}, {Sesar}, {Juri{\'c}},
  {Bond}, {Dalcanton}, {Rockosi}, {Yanny}, {Newberg}, {Beers}, {Allende
  Prieto}, {Wilhelm}, {Lee}, {Sivarani}, {Norris}, {Bailer-Jones}, {Re
  Fiorentin}, {Schlegel}, {Uomoto}, {Lupton}, {Knapp}, {Gunn}, {Covey},
  {Smith}, {Miknaitis}, {Doi}, {Tanaka}, {Fukugita}, {Kent}, {Finkbeiner},
  {Munn}, {Pier}, {Quinn}, {Hawley}, {Anderson}, {Kiuchi}, {Chen}, {Bushong},
  {Sohi}, {Haggard}, {Kimball}, {Barentine}, {Brewington}, {Harvanek},
  {Kleinman}, {Krzesinski}, {Long}, {Nitta}, {Snedden}, {Lee}, {Harris},
  {Brinkmann}, {Schneider}, \& {York}}]{Ivezic08}
{Ivezi{\'c}} {\v Z}. {et~al.}, 2008, \apj, 684, 287

\bibitem[{{Jiang} \& {Binney}(2000)}]{Jiang00}
{Jiang} I.-G., {Binney} J., 2000, \mnras, 314, 468

\bibitem[{{Kalberla} {et~al}\mbox{.}(2007){Kalberla}, {Dedes}, {Kerp}, \&
  {Haud}}]{Kalberla07}
{Kalberla} P.~M.~W., {Dedes} L., {Kerp} J., {Haud} U., 2007, \aap, 469, 511

\bibitem[{{Kazantzidis} {et~al}\mbox{.}(2008){Kazantzidis}, {Bullock},
  {Zentner}, {Kravtsov}, \& {Moustakas}}]{Kazantzidis08}
{Kazantzidis} S., {Bullock} J.~S., {Zentner} A.~R., {Kravtsov} A.~V.,
  {Moustakas} L.~A., 2008, \apj, 688, 254

\bibitem[{{Koposov} {et~al}\mbox{.}(2012){Koposov}, {Belokurov}, {Evans},
  {Gilmore}, {Gieles}, {Irwin}, {Lewis}, {Niederste-Ostholt}, {Pe{\~n}arrubia},
  {Smith}, {Bizyaev}, {Malanushenko}, {Malanushenko}, {Schneider}, \&
  {Wyse}}]{Koposov12}
{Koposov} S.~E. {et~al.}, 2012, \apj, 750, 80

\bibitem[{{Lake}(1989)}]{lake1989}
{Lake} G., 1989, \aj, 98, 1554

\bibitem[{{Laporte} {et~al}\mbox{.}(2016){Laporte}, {G{\'o}mez}, {Besla},
  {Johnston}, \& {Garavito-Camargo}}]{laporte2016}
{Laporte} C.~F.~P., {G{\'o}mez} F.~A., {Besla} G., {Johnston} K.~V.,
  {Garavito-Camargo} N., 2016, ArXiv e-prints

\bibitem[{{Li} {et~al}\mbox{.}(2012){Li}, {Newberg}, {Carlin}, {Deng}, {Newby},
  {Willett}, {Xu}, \& {Luo}}]{Li12}
{Li} J., {Newberg} H.~J., {Carlin} J.~L., {Deng} L., {Newby} M., {Willett}
  B.~A., {Xu} Y., {Luo} Z., 2012, \apj, 757, 151

\bibitem[{{Lindegren} {et~al}\mbox{.}(2016){Lindegren}, {Lammers}, {Bastian},
  {Hern{\'a}ndez}, {Klioner}, {Hobbs}, {Bombrun}, {Michalik}, {Ramos-Lerate},
  {Butkevich}, {Comoretto}, {Joliet}, {Holl}, {Hutton}, {Parsons},
  {Steidelm{\"u}ller}, {Abbas}, {Altmann}, {Andrei}, {Anton}, {Bach},
  {Barache}, {Becciani}, {Berthier}, {Bianchi}, {Biermann}, {Bouquillon},
  {Bourda}, {Br{\"u}semeister}, {Bucciarelli}, {Busonero}, {Carlucci},
  {Casta{\~n}eda}, {Charlot}, {Clotet}, {Crosta}, {Davidson}, {de Felice},
  {Drimmel}, {Fabricius}, {Fienga}, {Figueras}, {Fraile}, {Gai}, {Garralda},
  {Geyer}, {Gonz{\'a}lez-Vidal}, {Guerra}, {Hambly}, {Hauser}, {Jordan},
  {Lattanzi}, {Lenhardt}, {Liao}, {L{\"o}ffler}, {McMillan}, {Mignard}, {Mora},
  {Morbidelli}, {Portell}, {Riva}, {Sarasso}, {Serraller}, {Siddiqui}, {Smart},
  {Spagna}, {Stampa}, {Steele}, {Taris}, {Torra}, {van Reeven}, {Vecchiato},
  {Zschocke}, {de Bruijne}, {Gracia}, {Raison}, {Lister}, {Marchant},
  {Messineo}, {Soffel}, {Osorio}, {de Torres}, \& {O'Mullane}}]{Lindegren16}
{Lindegren} L. {et~al.}, 2016, \aap, 595, A4

\bibitem[{{Majewski} {et~al}\mbox{.}(2004){Majewski}, {Ostheimer},
  {Rocha-Pinto}, {Patterson}, {Guhathakurta}, \& {Reitzel}}]{majewski2004}
{Majewski} S.~R., {Ostheimer} J.~C., {Rocha-Pinto} H.~J., {Patterson} R.~J.,
  {Guhathakurta} P., {Reitzel} D., 2004, \apj, 615, 738

\bibitem[{{Martin} {et~al}\mbox{.}(2004{\natexlab{a}}){Martin}, {Ibata},
  {Bellazzini}, {Irwin}, {Lewis}, \& {Dehnen}}]{martin2004a}
{Martin} N.~F., {Ibata} R.~A., {Bellazzini} M., {Irwin} M.~J., {Lewis} G.~F.,
  {Dehnen} W., 2004{\natexlab{a}}, \mnras, 348, 12

\bibitem[{{Martin} {et~al}\mbox{.}(2005){Martin}, {Ibata}, {Conn}, {Lewis},
  {Bellazzini}, \& {Irwin}}]{martin2005}
{Martin} N.~F., {Ibata} R.~A., {Conn} B.~C., {Lewis} G.~F., {Bellazzini} M.,
  {Irwin} M.~J., 2005, \mnras, 362, 906

\bibitem[{{Martin} {et~al}\mbox{.}(2004{\natexlab{b}}){Martin}, {Ibata},
  {Conn}, {Lewis}, {Bellazzini}, {Irwin}, \& {McConnachie}}]{Martin04}
{Martin} N.~F., {Ibata} R.~A., {Conn} B.~C., {Lewis} G.~F., {Bellazzini} M.,
  {Irwin} M.~J., {McConnachie} A.~W., 2004{\natexlab{b}}, \mnras, 355, L33

\bibitem[{{Martin}, {Ibata} \& {Irwin}(2007){Martin}, {Ibata}, \&
  {Irwin}}]{martin2007}
{Martin} N.~F., {Ibata} R.~A., {Irwin} M., 2007, \apjl, 668, L123

\bibitem[{{Martin} {et~al}\mbox{.}(2014){Martin}, {Ibata}, {Rich}, {Collins},
  {Fardal}, {Irwin}, {Lewis}, {McConnachie}, {Babul}, {Bate}, {Chapman},
  {Conn}, {Crnojevi{\'c}}, {Ferguson}, {Mackey}, {Navarro}, {Pe{\~n}arrubia},
  {Tanvir}, \& {Valls-Gabaud}}]{martin2014}
{Martin} N.~F. {et~al.}, 2014, \apj, 787, 19

\bibitem[{{Martin} {et~al}\mbox{.}(2006){Martin}, {Irwin}, {Ibata}, {Conn},
  {Lewis}, {Bellazzini}, {Chapman}, \& {Tanvir}}]{martin2006}
{Martin} N.~F., {Irwin} M.~J., {Ibata} R.~A., {Conn} B.~C., {Lewis} G.~F.,
  {Bellazzini} M., {Chapman} S., {Tanvir} N., 2006, \mnras, 367, L69

\bibitem[{{Mateu} {et~al}\mbox{.}(2009){Mateu}, {Vivas}, {Zinn}, {Miller}, \&
  {Abad}}]{mateu2009}
{Mateu} C., {Vivas} A.~K., {Zinn} R., {Miller} L.~R., {Abad} C., 2009, \aj,
  137, 4412

\bibitem[{{Meisner} {et~al}\mbox{.}(2012){Meisner}, {Frebel}, {Juri{\'c}}, \&
  {Finkbeiner}}]{Meisner12}
{Meisner} A.~M., {Frebel} A., {Juri{\'c}} M., {Finkbeiner} D.~P., 2012, \apj,
  753, 116

\bibitem[{{Merrifield}(1992)}]{Merrifield92}
{Merrifield} M.~R., 1992, \aj, 103, 1552

\bibitem[{{Michel-Dansac} {et~al}\mbox{.}(2011){Michel-Dansac}, {Abadi},
  {Navarro}, \& {Steinmetz}}]{michel2011}
{Michel-Dansac} L., {Abadi} M.~G., {Navarro} J.~F., {Steinmetz} M., 2011,
  \mnras, 414, L1

\bibitem[{{Mihos} \& {Hernquist}(1994)}]{mihos1994}
{Mihos} J.~C., {Hernquist} L., 1994, \apjl, 425, L13

\bibitem[{{Momany} {et~al}\mbox{.}(2004){Momany}, {Zaggia}, {Bonifacio},
  {Piotto}, {De Angeli}, {Bedin}, \& {Carraro}}]{Momany04}
{Momany} Y., {Zaggia} S.~R., {Bonifacio} P., {Piotto} G., {De Angeli} F.,
  {Bedin} L.~R., {Carraro} G., 2004, \aap, 421, L29

\bibitem[{{Newberg} {et~al}\mbox{.}(2002){Newberg}, {Yanny}, {Rockosi},
  {Grebel}, {Rix}, {Brinkmann}, {Csabai}, {Hennessy}, {Hindsley}, {Ibata},
  {Ivezi{\'c}}, {Lamb}, {Nash}, {Odenkirchen}, {Rave}, {Schneider}, {Smith},
  {Stolte}, \& {York}}]{Newberg02}
{Newberg} H.~J. {et~al.}, 2002, \apj, 569, 245

\bibitem[{{P{\^a}ris} {et~al}\mbox{.}(2017){P{\^a}ris}, {Petitjean}, {Ross},
  {Myers}, {Aubourg}, {Streblyanska}, {Bailey}, {Armengaud},
  {Palanque-Delabrouille}, {Y{\`e}che}, {Hamann}, {Strauss}, {Albareti},
  {Bovy}, {Bizyaev}, {Niel Brandt}, {Brusa}, {Buchner}, {Comparat}, {Croft},
  {Dwelly}, {Fan}, {Font-Ribera}, {Ge}, {Georgakakis}, {Hall}, {Jiang},
  {Kinemuchi}, {Malanushenko}, {Malanushenko}, {McMahon}, {Menzel}, {Merloni},
  {Nandra}, {Noterdaeme}, {Oravetz}, {Pan}, {Pieri}, {Prada}, {Salvato},
  {Schlegel}, {Schneider}, {Simmons}, {Viel}, {Weinberg}, \& {Zhu}}]{Paris17}
{P{\^a}ris} I. {et~al.}, 2017, \aap, 597, A79

\bibitem[{{Pe{\~n}arrubia} {et~al}\mbox{.}(2005){Pe{\~n}arrubia},
  {Mart{\'{\i}}nez-Delgado}, {Rix}, {G{\'o}mez-Flechoso}, {Munn}, {Newberg},
  {Bell}, {Yanny}, {Zucker}, \& {Grebel}}]{Penarrubia05}
{Pe{\~n}arrubia} J. {et~al.}, 2005, \apj, 626, 128

\bibitem[{{Pier} {et~al}\mbox{.}(2003){Pier}, {Munn}, {Hindsley}, {Hennessy},
  {Kent}, {Lupton}, \& {Ivezi{\'c}}}]{Pier03}
{Pier} J.~R., {Munn} J.~A., {Hindsley} R.~B., {Hennessy} G.~S., {Kent} S.~M.,
  {Lupton} R.~H., {Ivezi{\'c}} {\v Z}., 2003, \aj, 125, 1559

\bibitem[{{Pillepich}, {Madau} \& {Mayer}(2015){Pillepich}, {Madau}, \&
  {Mayer}}]{pilepich2015}
{Pillepich} A., {Madau} P., {Mayer} L., 2015, \apj, 799, 184

\bibitem[{{Purcell} {et~al}\mbox{.}(2011){Purcell}, {Bullock}, {Tollerud},
  {Rocha}, \& {Chakrabarti}}]{Purcell11}
{Purcell} C.~W., {Bullock} J.~S., {Tollerud} E.~J., {Rocha} M., {Chakrabarti}
  S., 2011, \nat, 477, 301

\bibitem[{{Read} {et~al}\mbox{.}(2008){Read}, {Lake}, {Agertz}, \&
  {Debattista}}]{read2008}
{Read} J.~I., {Lake} G., {Agertz} O., {Debattista} V.~P., 2008, \mnras, 389,
  1041

\bibitem[{{Reid} \& {Brunthaler}(2004)}]{Reid04}
{Reid} M.~J., {Brunthaler} A., 2004, \apj, 616, 872

\bibitem[{{Rocha-Pinto} {et~al}\mbox{.}(2003){Rocha-Pinto}, {Majewski},
  {Skrutskie}, \& {Crane}}]{rocha2003}
{Rocha-Pinto} H.~J., {Majewski} S.~R., {Skrutskie} M.~F., {Crane} J.~D., 2003,
  \apjl, 594, L115

\bibitem[{{Rocha-Pinto} {et~al}\mbox{.}(2004){Rocha-Pinto}, {Majewski},
  {Skrutskie}, {Crane}, \& {Patterson}}]{rocha2004}
{Rocha-Pinto} H.~J., {Majewski} S.~R., {Skrutskie} M.~F., {Crane} J.~D.,
  {Patterson} R.~J., 2004, \apj, 615, 732

\bibitem[{{Sch{\"o}nrich}, {Binney} \& {Dehnen}(2010){Sch{\"o}nrich}, {Binney},
  \& {Dehnen}}]{Schonrich10}
{Sch{\"o}nrich} R., {Binney} J., {Dehnen} W., 2010, \mnras, 403, 1829

\bibitem[{{Sharma} {et~al}\mbox{.}(2011){Sharma}, {Bland-Hawthorn}, {Johnston},
  \& {Binney}}]{Sharma11}
{Sharma} S., {Bland-Hawthorn} J., {Johnston} K.~V., {Binney} J., 2011, \apj,
  730, 3

\bibitem[{{Sheffield} {et~al}\mbox{.}(2014){Sheffield}, {Johnston}, {Majewski},
  {Damke}, {Richardson}, {Beaton}, \& {Rocha-Pinto}}]{Sheffield14}
{Sheffield} A.~A., {Johnston} K.~V., {Majewski} S.~R., {Damke} G., {Richardson}
  W., {Beaton} R., {Rocha-Pinto} H.~J., 2014, \apj, 793, 62

\bibitem[{{Slater} {et~al}\mbox{.}(2014){Slater}, {Bell}, {Schlafly},
  {Morganson}, {Martin}, {Rix}, {Pe{\~n}arrubia}, {Bernard}, {Ferguson},
  {Martinez-Delgado}, {Wyse}, {Burgett}, {Chambers}, {Draper}, {Hodapp},
  {Kaiser}, {Magnier}, {Metcalfe}, {Price}, {Tonry}, {Wainscoat}, \&
  {Waters}}]{Slater14}
{Slater} C.~T. {et~al.}, 2014, \apj, 791, 9

\bibitem[{{Steinmetz} \& {Navarro}(2002)}]{steinmetz2002}
{Steinmetz} M., {Navarro} J.~F., 2002, \na, 7, 155

\bibitem[{{Toomre}(1977)}]{toomre1977}
{Toomre} A., 1977, \araa, 15, 437

\bibitem[{{Velazquez} \& {White}(1999)}]{velazquez1999}
{Velazquez} H., {White} S.~D.~M., 1999, \mnras, 304, 254

\bibitem[{{Wouterloot} {et~al}\mbox{.}(1990){Wouterloot}, {Brand}, {Burton}, \&
  {Kwee}}]{Wouterloot90}
{Wouterloot} J.~G.~A., {Brand} J., {Burton} W.~B., {Kwee} K.~K., 1990, \aap,
  230, 21

\bibitem[{{Xu} {et~al}\mbox{.}(2015){Xu}, {Newberg}, {Carlin}, {Liu}, {Deng},
  {Li}, {Sch{\"o}nrich}, \& {Yanny}}]{Xu15}
{Xu} Y., {Newberg} H.~J., {Carlin} J.~L., {Liu} C., {Deng} L., {Li} J.,
  {Sch{\"o}nrich} R., {Yanny} B., 2015, \apj, 801, 105

\bibitem[{{Xue} {et~al}\mbox{.}(2015){Xue}, {Rix}, {Ma}, {Morrison}, {Bovy},
  {Sesar}, \& {Janesh}}]{xue2015}
{Xue} X.-X., {Rix} H.-W., {Ma} Z., {Morrison} H., {Bovy} J., {Sesar} B.,
  {Janesh} W., 2015, \apj, 809, 144

\bibitem[{{Yanny} {et~al}\mbox{.}(2003){Yanny}, {Newberg}, {Grebel}, {Kent},
  {Odenkirchen}, {Rockosi}, {Schlegel}, {Subbarao}, {Brinkmann}, {Fukugita},
  {Ivezic}, {Lamb}, {Schneider}, \& {York}}]{Yanny03}
{Yanny} B. {et~al.}, 2003, \apj, 588, 824

\bibitem[{{Younger} {et~al}\mbox{.}(2008){Younger}, {Besla}, {Cox},
  {Hernquist}, {Robertson}, \& {Willman}}]{younger2008}
{Younger} J.~D., {Besla} G., {Cox} T.~J., {Hernquist} L., {Robertson} B.,
  {Willman} B., 2008, \apjl, 676, L21

\bibitem[{{Zacharias}, {Rafferty} \& {Zacharias}(2000){Zacharias}, {Rafferty},
  \& {Zacharias}}]{Zacharias00}
{Zacharias} N., {Rafferty} T.~J., {Zacharias} M.~I., 2000, in Astronomical
  Society of the Pacific Conference Series, Vol. 216, Astronomical Data
  Analysis Software and Systems IX, {Manset} N., {Veillet} C., {Crabtree} D.,
  eds., p. 427

\end{thebibliography}

\label{lastpage}

\end{document}